\documentstyle[prl,aps,twocolumn,psfig]{revtex}

\draft
\begin{document}

\twocolumn[\hsize\textwidth\columnwidth\hsize\csname @twocolumnfalse\endcsname
\title{Static and dynamic properties of ferroelectric thin film multilayers}
\author{M.D. Glinchuk and E.A. Eliseev}
\address{Institute for Materials Sciences, NASC of\\
Ukraine, Krjijanovskogo 3, 252180, Kiev, Ukraine}
\author{V. A. Stephanovich}
\address{Institute of Physics of Semiconductors, NASC of Ukraine, Prosp.\ Nauki 45,\\
252650, Kiev, Ukraine}
\author{M. G. Karkut and R. Farhi}
\address{Laboratory of Condensed Matter Physics, University of Picardy, 33 rue\\
Saint-Leu, 80039, Amiens, France}
\date{\today}
\maketitle

\begin{abstract}
A thermodynamic theory of ferroelectric thin film multilayers is developed.
The free energy functional is written down using a multilayer model in which
c-domain layers of one ferroelectric material alternate with a-domain layers
of a second ferroelectric material. We assume that the interfaces are
perfectly sharp and that the polarization at these boundaries is zero. The
renormalization of the free energy coefficients due to the stresses in the
films and to the depolarizing field was taken into account, as well as the
renormalization of the coefficients of the polarization gradients. The
equilibrium inhomogeneous polarization temperature and its thickness
dependence were determined from the solutions of the Euler-Lagrange
equations resulting from the minimization of the free energy functional. A
thickness induced ferroelectric phase transition is shown to exist and its
transition temperature and critical layer thickness depend on the domain
orientation. The criteria for ''a/c'', ''c/c'' and ''a/a'' domain
multilayering are calculated and expressed via coefficients of the free
energy density and the layer thickness.

The differential equation to calculate the inhomogeneous dielectric
susceptibility was found to be a Lam\'{e} equation. The calculated
multilayer susceptibility diverges at the transition temperature of the
thickness induced ferroelectric phase. This divergence is shown to be the
origin of the giant dielectric response observed in some multilayers. The
theory gives an excellent fit to the temperature dependence of the giant
susceptibility observed recently in multilayers of PbTiO$_{3}$-Pb$_{1-x}$La$_{x}$TiO$_{3}$ ($x=0.28$). The equation determining the time and space
dependence of the polarization is proposed and solved. The calculated
dispersion law for the nonlinear polarization waves in multilayer structures
reveals a critical wave vector for a given layer thickness, or a critical
layer thickness for a given wave vector, for which the frequency $\omega =0$. The frequency of the polarization waves is also shown to increase with
layer thickness.
\end{abstract}
\pacs{PACS numbers: 6865.+g, 7755.+f, 7784.-s}]

\section{Introduction}

Artificial thin film multilayers composed of alternating layers of different
materials have been the subject of study for many years. Much attention has
been devoted to semiconducting \cite{1}, metallic \cite{2}, magnetic \cite{3}
and superconducting \cite{4} multilayer structures. Increasingly frequent
investigations of ferroelectric oxide multilayers are now taking place due
to their technological promise: the artificial modification of structural
and physical properties for use in dielectric capacitors, memory systems,
pyroelectric detection and other types of devices makes them technologically
attractive (see e.g. \cite{sw,tab} and references therein), and there is
also a basic interest in creating model structures for which to study
fundamental questions related to ferroelectricity on increasingly smaller
length scales. Recent experiments have revealed many unusual properties,
such as giant dielectric response \cite{5}, anomalies in the
ferroelectric-paraelectric phase transition temperature \cite{6}, in
superlattice growth \cite{7}, and in superlattice structural anomalies \cite
{8}. In the latter work, investigation of PbTiO$_{3}$/BaTiO$_{3}$ (PT/BT)
multilayers with bilayer PT/BT thickness between $50A^{o}<\Lambda <360A^{o}$
\cite{8} shows that ''c''-domain BT layers alternate with ''a''-domain PT
layers. This is in contrast both to the ''c''-domain structure exhibited by
thin single PT films of the same thickness grown on the same substrate.
Raman spectra measurements reinforce this x-ray determined orientational
anomaly in addition to revealing the existence of a superlattice wavelength
dependent mode, whose frequency increases with increasing $\Lambda $. Up to
now there have been few theoretical studies of ferroelectric multilayer
structures. Calculation of the dielectric response has shown \cite{9} that
the spatial distribution of the layer thickness can lead to enhanced
dielectric properties over a broad temperature range. A phenomenological
theory has recently been developed \cite{qu} for thin film multilayers. The
spontaneous polarization and the dielectric susceptibility were calculated
numerically with parameters appropriate to PT/BT.\ They showed that a
thickness induced ferroelectric phase transition occurs only when the
strength of the interfacial coupling is weak.

In this work we will\ consider a multilayer built up from alternating (100)
and (001) ferroelectric thin films (such as PbTiO$_{3}$ and BaTiO$_{3}$)
epitaxially grown in the cubic paraelectric phase onto a cubic (001)
substrate. We use a thermodynamic approach to describe the static and
dynamic multilayer properties. The static properties are considered in
section II in which section IIA is devoted to the description of the free
energy of a multilayer with contributions coming from the internal stresses
and the depolarizing field. Calculations of the inhomogeneous polarization
and the thickness induced ferroelectric phase transition are performed in
IIB. The criterion for the presence of ''a/c'', ''a/a'' or ''c/c'' layer
structures in the superlattice is determined in IIC. The differential
equation giving the static dielectric susceptibility of multilayers is
considered in IID. The dynamic properties are treated in Section III. In
IIIA the differential equation governing the time and space dependence of
the polarization is written down and solved, and in IIIB the dispersion law
for nonlinear waves formed by the polarization in the multilayer is
reported. In the Discussion we show that the theory is in good agreement
with the observed giant dielectric response in PLT/PT ferroelectric thin
film multilayers \cite{5} and in the Conclusion we discuss the applicability
of our results and their future development. The Appendices contain the
detailed calculations of (1) the criteria for ''a/c'', ''a/a'' and ''c/c''
layer structure of the superlattice, (2) the solution of the differential
equation for the inhomogeous dielectric susceptibility, (3) the multilayer
susceptibility, (4) the value of the susceptibility for the thick film
ferroelectric phase and for the thin film paraelectric phase, and (5) the
Curie-Weiss law in the vicinity of the thickness induced ferroelectric phase
transition.

\section{Static properties}

\subsection{Free energy}

We consider a multilayer (see Figure1) built up from ''A'' and ''B''
ferroelectric layers extending from $z=-L/2$ to $L/2$. Each layer has
thickness $l_{i}$ ($i=A,B$) so that the total multilayer thickness is $%
L=N(l_{A}+l_{B})$ where $\Lambda =(l_{A}+l_{B})$ is the multilayer
wavelength and $N$ is the number of wavelengths in the superlattice. The
free energy of the multilayer system can be written as

\begin{equation}
\Phi =\frac{1}{L}\int\limits_{-L/2}^{L/2}\left( {\Phi _{A}(z)+\Phi
_{B}(z)+\Phi _{A,B}(z)}\right) dz  \eqnum{1}
\end{equation}
where $\Phi _{A}$and $\Phi _{B}$ are the free energy densities of the $A$
layers and the $B$ layers respectively and $\Phi _{A,B}$ is the free energy
density resulting from the interaction between the layers.

We will start with the following forms for $\Phi _{A}$ and $\Phi _{B}$,
which can be obtained from those free energies having cubic symmetry, and
which allows for the symmetry lowering that is related to the nonequivalence
of the $z$ and the $x$, $y$ polarization components in the film due to the
contribution of the mechanical strains and to the depolarizing field. These
effects are incorporated into the renormalized coefficients.

\begin{eqnarray}
&&\Phi _{A}
=a_{3}P_{Az}^{2}+a_{1}(P_{Ax}^{2}+P_{Ay}^{2})+a_{11}(P_{Ax}^{4}+P_{Ay}^{4})+
\nonumber \\
&&+a_{33}P_{Az}^{4}+a_{13}P_{Az}^{2}(P_{Ax}^{2}+P_{Ay}^{2})+\alpha
_{33}\left(\frac{\partial P_{Az}}{\partial z}\right) ^{2}+  \nonumber \\
&&+\alpha _{11}\left[ \left( \frac{\partial P_{Ax}}{\partial x}%
\right) ^2+\left(\frac{\partial P_{Ay}}{\partial y} \right) ^2 \right] +  \eqnum{2a}  \label{2a} \\
&&+\alpha _{44}\left[ \left(\frac {\partial P_{Ax}}{\partial z}\right)%
 ^{2}+\left(\frac{\partial P_{Ay}}{\partial z}\right) ^{2}\right] ,
\nonumber
\end{eqnarray}

\begin{eqnarray}
&&\Phi _{B}
=b_{3}P_{Bz}^{2}+b_{1}(P_{Bx}^{2}+P_{By}^{2})+b_{11}(P_{Bx}^{4}+P_{By}^{4})+
\nonumber \\
&&+b_{33}P_{Bz}^{4}+b_{13}P_{Bz}^{2}(P_{Bx}^{2}+P_{By}^{2})+\beta
_{33}\left( \frac{\partial P_{Bz}}{\partial z}\right) ^{2}+  \nonumber \\
&&+\beta _{11}\left[\left(\frac{ \partial P_{Bx}}{ \partial x}\right) ^{2}
+\left(\frac{\partial P_{By}}{\partial y}\right) ^2 \right] +  \eqnum{2b}  \label{2b} \\
&&+\beta _{44}\left[ \left( \frac {\partial P_{Bx}}{\partial z}\right) ^{2}+
\left(
\frac{\partial P_{By}}{\partial z}\right) ^{2}\right] .  \nonumber
\end{eqnarray}
Since we wish to take into account the change of $P_{x,y}$ along the $z$
direction, we have incorporated the additional gradient terms
$\partial P_{x,y}/\partial z$ into Eqs.(2).

Equations (2a) and (2b) correspond to the general case in which the x,y, and
z polarization components exist in both the A and in the B layers.
Experimental data for multilayers usually show that c-domain layers
alternate with c-domain layers \cite{7}, but recent experimental results on
PbTiO$_{3}$/BaTiO$_{3}$ multilayers have shown \cite{8} that c-domain BT
layers alternate with $a$-domain PT layers in the multilayer structure. In
order to be able to treat this interesting latter case we will assume that
the A layers consist only of $c$-domains (i.e. $P_{A}=P_{Az}$, $%
P_{Ax}=P_{Ay}=0$) while the B layers consist only of $a$-domains (i.e. $%
P_{Bx}=P_{By}\neq 0$, $P_{Bz}=0$). The variation of the polarization in the
A and B layers in the vicinity of the $z=0$ boundary is schematically
depicted in Figure 2 in which the natural boundary condition for this
configuration is $P_{Az}(z=0)=P_{Bx}(z=0)=0$ . Since Fig. 2 represents one
modulation period of the superlattice, this same boundary condition must be
valid for all the interfaces that make up the multilayer structure, namely
\begin{equation}
P_{Az}(z_{j})=P_{Bx}(z_{j})=0,  \eqnum{3}  \label{3}
\end{equation}
\begin{eqnarray*}
z_{j} &=&-\frac{L}{2}+2jl_{A,B};\text{ }z_{j}=-\frac{L}{2}+(2j+1)l_{A,B}; \\
j &=&0,1,2...N-1.
\end{eqnarray*}
The superlattice periodicity also implies the following periodic
polarization condition,
\begin{eqnarray}
P_{Az}(z+j(l_{A}+l_{B})) &=&P_{Az}(z)  \nonumber \\
P_{Bx}(z+j(l_{A}+l_{B})) &=&P_{Bx}(z)  \eqnum{4}  \label{4}
\end{eqnarray}
Note that even in the case of a ''c-domain/c-domain'' or a
''a-domain/a-domain''\ multilayer structure, zero polarization at the
boundaries would be the most probable since the situation at the interfaces
is physically similar to that of domain boundaries in bulk ferroelectrics.

We will, in this work, neglect the interaction energy term $\Phi _{AB}$ due
to the polarization interaction between the layers since this interaction
decays rapidly, and because of the boundary conditions, we will also neglect
any surface energy contributions. Hence we can rewrite the free energy
density (2) equations as
\begin{equation}
f_{A}=a_{3}P_{Az}^{2}+a_{33}P_{Az}^{4}+\alpha _{33}\left( {\frac{dP_{Az}}{dz}%
}\right) ^{2}  \eqnum{5a}  \label{5a}
\end{equation}
\begin{equation}
f_{B}=2b_{1}P_{Bx}^{2}+2b_{11}P_{Bx}^{4}+2\beta _{44}\left( {\frac{dP_{Bx}}{%
dz}}\right) ^{2}  \eqnum{5b}  \label{5b}
\end{equation}

In Eqs.(5) we have conserved what is most essential for the analysis of thin
film multilayers: the gradient terms which take into account the
polarization change along the $z$ direction (see Fig.2). Thus, in our model,
the free energy density $f(z)=f_{A}(z)+f_{B}(z)$ will replace $\Phi
_{A}(z)+\Phi _{B}(z)+\Phi _{AB}(z)$. The most important feature of the free
energy density of thin films is its dependence on $z$. In the most general
case both the polarization $P$ and the mechanical stresses $\sigma $ can be
inhomogeneous so that their gradients $P^{\prime }$ and $\sigma ^{\prime }$%
should also be taken into account. Because of this, the free energy $\Phi $
is a functional of $P$, $P^{\prime }$, $\sigma $, $\sigma ^{\prime }$, which
we will write as:
\begin{equation}
\Phi (P,\sigma )=\int\limits_{-L/2}^{L/2}f\left( {P(z),P^{\prime }(z),\sigma
(z),\sigma ^{\prime }(z)}\right) dz  \eqnum{6}  \label{6}
\end{equation}
where, for the sake of clarity, we have omitted the vector and tensor
component notation.

The equilibrium values of polarization and stresses or strains must then
satisfy the Euler-Lagrange equations \cite{10}
\begin{equation}
\frac{\partial f}{\partial P}-\frac{d}{dz}\frac{\partial f}{\partial P^{\prime }}=0  \eqnum{7a}  \label{7a}
\end{equation}
\begin{equation}
\frac{\partial f}{\partial \sigma }-\frac{d}{dz}\frac{\partial f}%
{\partial \sigma ^{\prime }}=0  \eqnum{7b}
\end{equation}
with the corresponding boundary conditions for $P$, $\sigma $, $P^{\prime }$%
, $\sigma ^{\prime }$. Note that Eqs. (7a) and (7b) can be written for all
components of $P_{i}$ and $\sigma _{ij}$ respectively.

In what follows we will calculate $P(z)$ on the basis of Eq.(7a) and $\sigma
$ from first term in Eq.(7b) because we will here make the simplifying
assumption that the interfacial stresses are homogeneous and so we will
neglect the second term in Eq. (7b). For films, the coefficients of Eq. (2)
are renormalized coefficients. The stresses as well as the depolarizing
field act on the coefficients of the free energy of cubic symmetry. In
general the thickness of the substrate is much larger than that of the
multilayer structure. Taking this into consideration for the equilibrium
condition of the mechanical forces, it can be shown \cite{11} that the
stresses induced by upper layers in the underlying films are negligibly
small. Thus the nonzero stresses are $\sigma _{xx}=\sigma _{yy}$ and $\sigma
_{xy}=0$ while $\sigma _{zz}=\sigma _{xz}=\sigma _{yz}=0$ because of the
existence of the unstressed free surface on top of the multilayer. The
renormalization of the free energy coefficients $a_{i}$, $a_{ij}$, ($i$, $j$%
, $k$ - Voigt notation) of bulk cubic symmetry by these stresses was
previously performed \cite{12,13} for a single thin film on a cubic
substrate. Using the results of \cite{13} along with the depolarizing field
contribution, the coefficients $a_{3}$ and $a_{1}$ in Eq.(2a) can be written
as :
\begin{equation}
a_{3}=a-X\frac{2Q_{12}^{A}}{S_{11}^{A}+S_{12}^{A}}+\frac{2\pi }{\varepsilon
_{A}}  \eqnum{8a}  \label{8a}
\end{equation}
\begin{equation}
a_{1}=a-X\frac{Q_{11}^{A}+Q_{12}^{A}}{S_{11}^{A}+S_{12}^{A}}  \eqnum{8b}
\label{8b}
\end{equation}
Here $a=a_{0}^{A}(T-T_{c0}^{A})$, $T_{c0}^{A}$, $Q_{ij}^{A}$ and $S_{ij}^{A}$
are respectively the coefficient of the free energy, the ferroelectric phase
transition temperature, the electrostriction constants and the elastic
modulus of the A material. The strain $X=X_{1}=X_{2}$ can be represented as
\begin{equation}
X=X_{mf}+X_{th}+X_{dis}  \eqnum{8c}  \label{8c}
\end{equation}
where $X_{mf}=(b-a)/a$, $X_{th}=(\alpha _{B}-\alpha _{A})(T-T_{g})$ and $%
X_{dis}$ are respectively the misfit, thermal and disorder induced strains.
In these expressions, $b$ and $a$ are the cubic lattice constants, $\alpha
_{A}$ and $\alpha _{B}$ are the thermal expansion coefficients of the two
materials in the cubic bulk phase and $T_{g}$ is the growth temperature. We
point out that the relaxation processes related to misfit dislocations,
domain structure appearance and impurity diffusion processes can decrease $X$%
, but full relaxation can only be achieved in bulk materials \cite{14}. The
last term in Eq.(8a) originates from the depolarizing field $E_{d}$ whose
contribution equals to $-1/2E_{d}P_{Az}$ with $E_{d}=-4\pi
P_{Az}/\varepsilon _{A}$ for a free standing film without electrodes ($%
\varepsilon _{A}$ is\ the dielectric constant of the A material \cite{15}).
The renormalized coefficients in Eq.(2b) can be simply obtained from Eqs.(8)
by the substitution $a_{3}\rightarrow b_{3}$, $a_{1}\rightarrow b_{1}$ and
the index A $\rightarrow $ B.

The renormalization of the transition temperature in the layer follows from
Eqs.(8) and its analog for the B layer. For the A and B layers the
transition temperatures can be written as:
\begin{equation}
T_{cz}^{A,B}=T_{c0}^{A,B}+\frac{X^{A,B}2Q_{12}^{A,B}}{%
a_{0}^{A,B}(S_{11}^{A,B}+S_{12}^{A,B})}-\frac{2\pi }{a_{0}^{A,B}\varepsilon
_{A,B}}  \eqnum{9a}  \label{9a}
\end{equation}
\begin{equation}
T_{cx}^{A,B}=T_{c0}^{A,B}+\frac{X^{A,B}}{a_{0}^{A,B}}\frac{%
(Q_{11}^{A,B}+Q_{12}^{A,B})}{(S_{11}^{A,B}+S_{12}^{A,B})}  \eqnum{9b}
\label{9b}
\end{equation}
where $T_{cz}$ and $T_{cx}$ are respectively the transition temperatures for
the appearance of $P_{z}\neq 0$ ($c$-domain structure) and $P_{x}\neq 0$ ($a$%
-domain structure). It is seen that the depolarizing field contribution
decreases the transition temperature of the $c$-domain layer whereas the
influence of strain on this temperature depends on the signs of the
electrostriction constants and whether the strains are tensile ($X>0$) or
compressive ($X<0$) \cite{12}.

We have omitted terms to the sixth power and above in the polarization,
which is the correct procedure only for second order ferroelectric phase
transitions. In general, bulk ferroelectric materials such as BaTiO$_{3}$
and PbTiO$_{3}$ undergo first order phase transitions, but calculations on
BaTiO$_{3}$ and PbTiO$_{3}$ thin films have shown \cite{13} that the
renormalized coefficients of $P_{z}^{4}$ and $P_{x}^{4}$ have positive
values and thus the ferroelectric thin film transition is second order
rather than first order. For this reason we consider the free energy forms
(2) and (5) to be those appropriate for multilayer films.

To the best of our knowledge there has been no consideration of how the
stresses will renormalize the coefficients of the gradient terms in the free
energy of cubic symmetry. We have performed calculations similar to those in
\cite{13}, for which we have added the additional term $\Delta F$ to the
cubic symmetry free energy contained in Ref.16:
\begin{eqnarray}
\Delta F &=&\gamma \left[ {\ \left( {\frac{\partial P_{z}}{\partial z}}%
\right) ^{2}+\left( {%
\ \frac{\partial P_{x}}{\partial x}}\right) ^{2}+\left( {\ %
\frac{\partial P_{y}}{\partial y}}\right) ^{2}}%
\right] -  \nonumber \\
&&-\Biggl\{\delta {_{111}\left[ {\ \left( {\ \frac{\partial P_{x}}{\partial x}}\right)
^{2}+\left( {\frac{\partial P_{y}}{\partial y}}\right) ^{2}}\right] +}  \eqnum{10}
\label{10} \\
&&{+\delta _{133}\left( {\frac{\partial P_{z}}{\partial z}}\right) ^{2}+\delta _{144}\left[ {%
\left( {\frac{\partial P_{x}}{\partial z}}\right) ^{2}+%
\left( {\ \frac{\partial P_{y}}{\partial z}}\right)
^{2}}\right] }\Biggr\}\sigma _{1},  \nonumber
\end{eqnarray}
where $\sigma _{1}=\sigma _{xx}=\sigma _{yy}\neq 0$ is the nonzero
homogeneous stress in the film. Minimization of the free energy with respect
to $\sigma _{1}$ gives the renormalized coefficients:
\begin{eqnarray}
\alpha _{33} &=&\gamma -\frac{X\delta _{133}}{S_{11}+S_{{12}}}  \nonumber \\
\alpha _{11} &=&\gamma -\frac{X\delta _{111}}{S_{11}+S_{{12}}}  \eqnum{11}
\label{11} \\
\alpha _{44} &=&-\frac{X\delta _{144}}{S_{11}+S_{{12}}}  \nonumber
\end{eqnarray}
Obviously, the same relations hold for the free energy and the $\beta $
coefficients of the B material. The parameters $\gamma $, $S_{ij}$, $\delta
_{ijk}$ are material parameters. The coefficient $\alpha _{44}$ is
proportional to $X$, i.e. $\alpha _{44}\neq 0$ in films having nonzero
homogeneous strain $X$ (8c). The coefficients $\delta _{ijk}$ are components
of a sixth rank tensor which is the lowest rank (and so having the largest
components) which will relate the squared gradient terms and the components
of the stress tensor.

\subsection{Polarization}

The Euler-Lagrange equation (7a) with $f=f_{A}+f_{B}$ (see Eqs.(5)) for $%
P=P_{Az}$ or $P_{Bx}$ and $P^{\prime }=(dP_{Az})/(dz)$ or $(dP_{Bx})/(dz)$
makes it possible to find the equilibrium values of the polarization on the
basis of the equations:
\begin{equation}
a_{3}P_{Az}+2a_{33}P_{Az}^{3}-\alpha _{33}\frac{d^{2}P_{Az}}{dz^{2}}=0
\eqnum{12a}  \label{12a}
\end{equation}
\begin{equation}
b_{1}P_{Bx}+2b_{11}P_{Bx}^{3}-\beta _{44}\frac{d^{2}P_{Bx}}{dz^{2}}=0
\eqnum{12b}  \label{12b}
\end{equation}

The above equations have to be solved subject to the periodicity conditions
(3) and (4). We shall demonstrate the solution for $P_{Az}$ only, since $%
P_{Bx}$ can be obtained from $P_{Az}$ by substituting the corresponding
coefficients. To integrate Eq.(12a) we let $(dP_{Az})/(dz)=g(P_{Az})$. This
gives
\begin{equation}
\frac{d^{2}P_{Az}}{dz^{2}}=g(P_{Az})\frac{dg(P_{Az})}{dP_{Az}}  \eqnum{13}
\label{13}
\end{equation}

Substitution of Eq.(13) into (12a) leads to
\begin{equation}
a_3P_{Az}+2a_{33}P_{Az}^3=\alpha _{33}g(P_{Az})\frac{dg(P_{Az})}{dP_{Az}}
\eqnum{14}  \label{14}
\end{equation}
which gives after integration
\begin{equation}
a_3P_{Az}^2+a_{33}P_{Az}^4=\alpha _{33}g^2(P_{Az})+c_{33}  \eqnum{15}
\label{15}
\end{equation}
To obtain the constant $c_{33}$ we introduce the maximum polarization in the
layer $P_{Azm}$ which satisfies the condition $\frac{dP_{Az}}{dz}\mid
_{P_{Az}=P_{Azm}}=0$, and we find
\begin{equation}
c_{33}=a_3P_{Azm}^2+a_{33}P_{Azm}^4  \eqnum{16}  \label{16}
\end{equation}

Substitution of Eq.(16) into Eq.(15) leads to
\begin{equation}
a_{3}(P_{Az}^{2}-P_{Azm}^{2})+a_{33}(P_{Az}^{4}-P_{Azm}^{4})=\alpha
_{33}\left( {\ \frac{dP_{Az}}{dz}}\right) ^{2}  \eqnum{17}  \label{17}
\end{equation}
We now introduce the following parametrization
\begin{equation}
P_{Az}(z)=P_{Azm}\sin \theta _{A}(z)  \eqnum{18}  \label{18}
\end{equation}
\begin{equation}
k_{Az}^{2}=\frac{P_{Azm}^{2}}{2P_{Az0}^{2}-P_{Azm}^{2}}  \eqnum{19}
\label{19}
\end{equation}
where $P_{Az0}^{2}=-a_{3}/(2a_{33})$ is the homogeneous polarization in a
thick film - when the derivative in Eq.(12a) can be neglected. Note that
this is {\it not} the polarization in the bulk material because the
parameters $a_{3}$ and $a_{33}$ have been renormalized by the stresses in
the layers (see Eqs.(8)). Because the homogeneous polarization corresponds
to the mean field approximation, $P_{Az0}$ can be considered as the film
polarization calculated in this approximation. Substituting Eqs. (18) and
(19) into (17) gives:
\begin{equation}
\alpha _{33}\left( {\frac{d\theta _{A}}{dz}}\right) ^{2}=-\frac{a_{3}}{%
1+k_{Az}^{2}}(1-k_{Az}^{2}\sin ^{2}\theta _{A})  \eqnum{20}  \label{20}
\end{equation}
After separating the variables we obtain

\begin{eqnarray*}
dz_{3} &=&\sqrt{1+k_{Az}^{2}}\frac{d\theta _{A}}{\sqrt{1-k_{Az}^{2}\sin
^{2}\theta _{A}}} \\
\int\limits_{-L_{z}/2+2jl_{Az}}^{z_{3}}dz_{3} &=&\sqrt{1+k_{Az}^{2}}%
\int\limits_{0}^{\theta _{A}}\frac{d\theta _{A}}{\sqrt{1-k_{Az}^{2}\sin
^{2}\theta _{A}}}
\end{eqnarray*}
or
\begin{equation}
\frac{z_{3}+L_{z}/2-2jl_{Az}}{\sqrt{1+k_{Az}^{2}}}=\int\limits_{0}^{\theta
_{A}}\frac{d\theta _{A}}{\sqrt{1-k_{Az}^{2}\sin ^{2}\theta _{A}}}  \eqnum{21}
\label{21}
\end{equation}
where we have introduced the dimensionless variables
\begin{equation}
z_{3}=\sqrt{-\frac{a_{3}}{\alpha _{33}}}z;\quad l_{Az}=\sqrt{-\frac{a_{3}}{%
\alpha _{33}}}l_{A};\quad L_{z}=\sqrt{-\frac{a_{3}}{\alpha _{33}}}L
\eqnum{22}  \label{22}
\end{equation}

It follows from the theory of elliptic functions \cite{16,17,18} that for
the relation (21) the function $\theta _{A}(z)$ has the form
\[
\theta _{A}(z_{3})={\rm am}\left[ {\ \frac{z_{3}+L_{z}/2-2jl_{Az}}{\sqrt{%
1+k_{Az}^{2}}},k_{Az}}\right]
\]

Hence from Eq.(18) we obtain the solution in terms of the elliptic sine
function ${\rm sn}$:
\begin{equation}
P_{Az}(z_{3})=P_{Azm}{\rm sn}\left( {\frac{z_{3}+L_{z}/2-2jl_{Az}}{\sqrt{%
1+k_{Az}^{2}}},k_{Az}}\right)  \eqnum{23}  \label{23}
\end{equation}
It can be seen that $P_{Az}(z)$ satisfies Eq.(12a) and the periodicity
conditions of Eqs.(4). $P_{Az}(z)$ is depicted graphically in Fig.3.

The relation between $P_{Azm}$ and $l_{Az}$ can be found from the
periodicity condition of the elliptic sine function (see \cite{16,17,18}),
namely
\begin{equation}
{\rm sn}\left[ {\ z_{3}=-\frac{{L_{z}}}{2}+(2j+1)l_{Az}}\right] =2K(k_{Az})
\eqnum{24}  \label{24}
\end{equation}
where
\[
K(k_{Az})=\int\limits_{0}^{\pi /2}\frac{d\theta }{\sqrt{1-k_{Az}^{2}\sin
^{2}\theta }}
\]
is the complete elliptic integral of the first kind.

Substitution of (24) into (23) gives
\begin{equation}
l_{Az}=2\sqrt{1+k_{Az}^{2}}K(k_{Az})  \eqnum{25}  \label{25}
\end{equation}
We plot the dependence Eq. (25) in Fig.4. Since we can write the
polarization ratio as $P_{Azm}^{2}/P_{Az0}^{2}=2k_{Az}^{2}/(1+k_{Az}^{2})$
(see Eq.(19)), Fig.4 makes it possible to obtain the dependence of this
ratio on the film thickness. We can see that for thick enough films (for the
dimensionless length $l_{Az}\approx 10-12$) $P_{Azm}\approx P_{Az0}$, and so
this polarazation ratio (see the additional scale in Fig.4) can be used as a
criterion for distinguishing between thick and thin films. Moreover, Fig. 4
shows the existence of a critical layer thickness $l_{Az}=\pi $ such that
the spontaneous polarization in the layer will exist only for $l_{Az}\geq
\pi $. Thus our calculations yield a thickness induced ferroelectric phase
transition which was previously discussed for multilayer films \cite{qu} and
for single ferroelectric thin films \cite{19,20}. The thickness induced
phase transition temperature $T_{cl}^{A}$ follows from $l_{Az}=\sqrt{%
-a_{0}^{A}(T-T_{cz}^{A})/\alpha _{33}}l_{A}=\pi $ :
\begin{equation}
T_{cl}^{A}=T_{cz}^{A}-\frac{\pi ^{2}}{l_{A}^{2}}\frac{\alpha _{33}}{a_{0}^{A}%
}  \eqnum{26a}  \label{26a}
\end{equation}
where $T_{cz}^{A}$, the renormalized layer temperature, is given by Eq(9a).
The reduced temperature $T_{cl}^{A}/T_{c0}^{A}$ as a function of reduced
thickness $l_{A}/l_{A0}$ is plotted in Fig.5. The characteristic thickness $%
l_{A}=l_{A0}$ at which $T_{cl}^{A}=0$ is:
\begin{equation}
l_{A0}=\pi \sqrt{\frac{\alpha _{33}}{a_{0}^{A}T_{cz}^{A}}}  \eqnum{26b}
\label{26b}
\end{equation}
The Fig. 5 plot is in agreement with available experimental data \cite{6} on
ferroelectric KNbO$_{3}$/KTaO$_{3}$ superlattices for intermediate
multilayer wavelengths. The range of the existence of the thickness induced
phase transition is given by
\begin{equation}
l_{A0}\leq l_{A}\leq \frac{\pi }{\sqrt{a_{0}^{A}(T_{cz}^{A}-T_{cl}^{A})/%
\alpha _{33}}}  \eqnum{26c}  \label{26c}
\end{equation}
with $l_{A}\rightarrow \infty $ at $T_{cl}^{A}\rightarrow T_{cz}^{A}$, the
renormalized transition temperature. Below a certain thickness, $%
l_{A}<l_{A0} $ there is no thickness induced phase transition because $%
T_{cl}^{A}$ becomes negative. The thickness dependence of $T_{cl}^{A}$ (see
Eq.(26a)) means that the thinner the film, the weaker its ferroelectricity.

The distribution of the polarization parallel to the multilayer growth axis,
$P_{Bx}(z)$ ,can be obtained from (23) by simply replacing the coefficients
(see (12a), (12b)):
\begin{equation}
a_{3}\rightarrow b_{1},\quad 2a_{33}\rightarrow b_{11},\quad \alpha
_{33}\rightarrow \beta _{33}  \eqnum{27}  \label{27}
\end{equation}
The solution for $P_{Bx}(z)$ thus has the form
\begin{equation}
P_{Bx}(z_{1})=P_{Bxm}{\rm sn}\left[ {\ \frac{z_{1}+L_{x}/2-(2j+1)l_{B}}{\sqrt{%
1+k_{Bx}^{2}}},k_{Bx}}\right]  \eqnum{28a}  \label{28a}
\end{equation}

\begin{equation}
k_{Bx}^{2}=\frac{P_{Bxm}^{2}}{2P_{Bx0}^{2}-P_{Bx}^{2}},\quad l_{Bx}=2\sqrt{%
1+k_{Bx}^{2}}K(k_{Bx})  \eqnum{28b}  \label{28b}
\end{equation}
where
\begin{equation}
z_{1}=\sqrt{-\frac{b_{1}}{\beta _{44}}}z,\quad l_{Bx}=\sqrt{-\frac{b_{1}}{%
\beta _{44}}}l_{B},\quad L_{x}=\sqrt{-\frac{b_{1}}{\beta _{44}}}L  \eqnum{29}
\label{29}
\end{equation}
Fig. 3 of course also represents $P_{Bx}(z)$.

The temperature of the thickness induced ferroelectric phase transition can
similarly be obtained from (26), (27) with the substitution $%
T_{cz}^{A}\rightarrow T_{cx}^{B}$, $a_{0}^{A}\rightarrow a_{0}^{B}$
\begin{equation}
T_{cl}^{B}=T_{cx}^{B}-\frac{\pi ^{2}}{l_{B}^{2}}\frac{\beta _{44}}{a_{0}^{B}}%
,\quad l_{B0}=\pi \sqrt{\frac{\beta _{44}}{a_{0}^{B}T_{cx}^{B}}}  \eqnum{30}
\label{30}
\end{equation}
Since the parameters in Eqs.(26) and (30) are different, the critical
characteristics of the thickness induced phase transition should also be
different in the A and the B layers. This may open up the prospect of
engineering new multilayer materials constructed with several ferroelectric
thin films (including superstructures consisting of several thin films in
its unit cell) with a broad distribution of the transition temperature. This
will result in a distribution of the material properties and also in any
anomalous behavior, which may then be exploitable for device applications.

\subsection{Criterion for ''a/c'', ''c/c'' and ''a/a'' domain structures}

Up until now we have considered a multilayer built up of layers in which the
polarization alternates between being in the plane of the film (a-domain
layers) and perpendicular to the plane of the film ( c-domain layers). This
choice was motivated by the experimental results of Ref\cite{8}. We will now
look at the conditions for which this situation is energetically favorable
compared to a- only \ or c- only layers throughout the multilayer structure
(a/a and c/c layering respectively). The requirement for the a/c multilayer
to occur is that the free energy should be less than that for c/c or a/a
multilayers. In supposing that the A layers have ''c''-domain structure, the
criterion for the existence of an a/c multilayer can be written as
\begin{equation}
F_{B}(P_{Bx})<F_{B}(P_{Bz})  \eqnum{31}  \label{31}
\end{equation}
where $F_{B}$ can be obtained by integration of the free energy density (5)
over $dz$ and summing over the layers. When the inequality (31) is not
satisfied we will have the ''c/c'' domain criterion. Eq.(31) will correspond
to ''a/a'' domain criterion if we suppose that the A layers are $a$-domain.
To calculate $F_{B}(P_{Bx})$ and $F_{B}(P_{Bz})$ we have to substitute
Eq.(23) into Eq.(5a) (with $A\rightarrow B$) and Eq.(28a) into Eq.(5b) and
then perform the integration over $dz$. This integration and summation (see
Appendix 1 for details) yield the following criterion for ''a/c'' domain
structure:
\begin{equation}
\frac{b_{3}^{2}}{b_{33}}\varphi (k_{Bz})<\frac{b_{1}^{2}}{b_{11}}\varphi
(k_{Bx})  \eqnum{32}  \label{32}
\end{equation}
Here
\begin{eqnarray}
\varphi (k_{Bi}) &=&\frac{1}{(1+k_{{Bi}}^{2})^{2}}\left[ {\ \frac{1}{2}%
k_{Bi}^{2}+1-f}(k_{Bi})\right] ,  \eqnum{33}  \label{33} \\
{f}(k_{Bi}) &=&{(1+k_{Bi}^{2})\frac{E(k_{Bi})}{K(k_{Bi})},\ \ }i=z,x,
\nonumber
\end{eqnarray}
where
\[
E(k)=\int\limits_{0}^{\pi /2}\sqrt{1-k^{2}\sin ^{2}\theta }d\theta
\]
is the complete elliptic integral of the second kind. The function $\varphi
(k)$ is calculated numerically and the results are shown in Figure 6. One
can see that $\varphi (k_{i})$ slowly increases for $0.3\leq k_{i}\leq 0.8$
and for $1\geq k_{i}>0.8$ it changes much faster. In the scale of
dimensionless thickness $l_{z}$ the region of slow increase corresponds to $%
5.5\geq l_{z,x}\geq 3.5$. In the thick film limit ($k_{Bi}=1$) $\varphi =3/8$
and the criterion (32) transforms into
\begin{equation}
\frac{b_{3}^{2}}{b_{33}}<\frac{b_{1}^{2}}{b_{11}}  \eqnum{34}  \label{34}
\end{equation}
This can be rewritten as
\begin{equation}
\frac{(T_{cz}^{B}-T)^{2}}{b_{33}}<\frac{(T_{cx}^{B}-T)^{2}}{b_{11}}
\eqnum{35}  \label{35}
\end{equation}
where the transition temperatures in the thick films $T_{cz,x}^{B}$ are
given by Eqs.(8). It is seen that for $T_{cx}^{B}>T_{cz}^{B}$ the preference
is for $a$-domain orientation in the B layers (and hence for a ''c/a''
domain multilayer). In comparing the formulas for $T_{cz}$ and $T_{cx}$
(Eqs. (9a) and (9b)), we see that for tensile strain ($X>0)$ and for $%
Q_{12}<0$ and $Q_{11}+Q_{12}>0$, which is the case for many ferroelectrics
with the perovskite structure, $T_{cx}$ will be higher than $T_{cz}$.
However for compressive strain ($X<0$) it is possible for $T_{cx}$ to be
smaller or larger than $T_{cz}$ depending on the depolarizing field. $%
T_{cx}^{B}>T_{cz}^{B}$ can result from the depolarization field's negative
contribution to $T_{cz}$. However, when the depolarization field can be
neglected (e.g. a film with electrodes having high conductivity \cite{15})
compressive strains will lead to $T_{cx}<T_{cz}$.

We point out that the criterion given by Eq.(35) corresponds to considering
the film in the mean field approximation with homogeneous polarization $%
P=P_{0}$ (see Eq.(19)). It is the function $\varphi $ which takes into
account the contribution of inhomogeneous polarization related to the
gradients in the free energy. Thus for the general case, one has to use
criterion Eq.(32) for ''a/c'' domains in the multilayer structure, keeping
in mind that the opposite condition (sign ''$>$'' substituted for sign ''$<$%
'' in Eq.(32)) implies the existence of a ''c/c''-domain multilayer
structure. The criterion for ''a/a'' domain coincides with Eq.(32) for the
case when the A layers are ''a''-domain. The criterion depends on the free
energy parameters, the transition temperatures, and the ratio of the maximum
polarization in a layer $P_{m}$ to the thick film polarization $P_{0}$. The
latter in turn depends on the film thickness, the coefficients of the
gradient terms and the free energy constants $b_{3}$, $b_{1}$. With the help
of Fig.4 or using the analytical formulas Eqs.(25) and (28b), Eq.(33) can be
rewritten in terms of a dimensionless layer thickness $l_{Bx,z}$ which is
presented as the second scale in Fig.6. By keeping in mind the analytical
form of $l_{Bx,z}$ (Eq.29) one can see that the criterion (Eq.32) depends on
the film thickness and the temperature dependent free energy parameters.
Knowledge of these parameters will make it possible to calculate completely
the criterion for a multilayer domain structure for a given layer thickness.

\subsection{Static dielectric susceptibility}

We will now consider a multilayer system in an external electric field $%
E=E_{z}$. In so doing we must add the term $-E_{z}P_{Az}$ to the free energy
density of the ferroelectric phase. From the general form of the
Euler-Lagrange equation (7a), it follows that we have to add $-E_{z}$ to
Eq.(12a). Therefore this equation now has the form:
\begin{equation}
2a_{3}P_{Az}+4a_{33}P_{Az}^{3}-E_{z}-2\alpha _{33}\frac{d^{2}P_{Az}}{dz^{2}}%
=0  \eqnum{36}  \label{36}
\end{equation}
The applied electric field induces an additional homogeneous polarization $%
\Delta P_{Az}=\chi _{zz}E_{z}$ where $\chi _{zz}$ is the linear dielectric
susceptibility ($E_{z}$ is assumed to be small). Therefore the polarization
in the A layer is now $P_{AzE}=P_{Az}+\Delta P_{Az}$ where $\Delta
P_{Az}<<P_{Az}$. Putting $P_{AzE}$ in Eq.(36) and keeping only terms to the
first power in $\Delta P_{Az}$ leads to the equation:

\begin{equation}
2\alpha _{33}\frac{d^{2}\chi _{zz}}{dz^{2}}-\chi _{zz}\left[
2a_{3}+12a_{33}P_{Az}^{2}\right] +1=0;\text{ }T\leq T_{cl}^{A}  \eqnum{37a}
\label{37}
\end{equation}
Eq.(37a) defines the static dielectric susceptibility for the case of
inhomogeneous polarization $P_{Az}(z)$. Homogeneous polarization $P_{Az0}$
leads to $\chi _{zz0}^{-1}=2a_{3}+12a_{33}P_{Az0}^{2}$. It is obvious that
the $\chi _{zz}(z)$ dependence originates from the z dependence of $%
P_{Az}(z) $. When the polarization is inhomogeneous, it is wrong to suppose
that the susceptibility in the ferroelectric phase can be found by
conventional differentiation of the free energy, i.e. $\chi
_{zz}^{-1}=2a_{3}+12a_{33}P_{Az}^{2}(z)$ is incorrect because $\chi _{zz}$
obviously does not satisfy Eq.(37a) in general.

We will now discuss the solutions of Eq.(37a). By introducing the
dimensionless variable $\qquad \qquad \qquad \qquad \qquad \qquad \qquad \xi
_{3}=\left( z_{3}+L_{z}/2-2jl_{Az}\right) /\sqrt{1+k_{Az}^{2}}$
and keeping in mind that $%
2a_{3}+12a_{33}P_{Az}^{2}(z)=2a_{3}(1+k_{Az}^{2}-6k_{Az}^{2}{\rm sn}^{2}(\xi
_{3}))/(1+k_{Az}^{2})$ we can rewrite Eq.(37a) in the form:
\begin{equation}
\frac{d^{2}\chi _{zz}}{d\xi _{3}^{2}}+\chi _{zz}\left(
1+k_{Az}^{2}-6k_{Az}^{2}{\rm sn}^{2}(\xi _{3})\right) =-2\left(
1+k_{Az}^{2}\right) \chi _{t}  \eqnum{37b}  \label{37b}
\end{equation}
where $\chi _{t}^{-1}=4\left| a_{3}\right| $ is the thick film
susceptibility. The homogeneous part of Eq.(37b) is known to be a Lam\'{e}
equation \cite{21}. The general solution of Eq.(37b) can be written as a sum
of the fundamental solutions of the homogeneous equation plus the particular
solution of the inhomogeneous equation. Letting $k_{Az}^{2}=m_{3}$ and
denoting the susceptibility at the boundaries as $\chi _{s}$ we obtain

\begin{figure}[th]
\vspace*{-10mm}
\centerline{\centerline{\psfig{figure=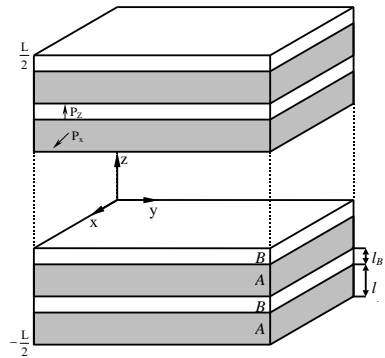,width=0.8\columnwidth}}}
\caption{Schematic diagram of a multilayer structure.}
\end{figure}

\twocolumn[\hsize\textwidth\columnwidth\hsize\csname @twocolumnfalse\endcsname
\begin{eqnarray}
\chi _{zz}(\xi _{3}) &=&\chi _{s}{\rm cn}(\xi _{3})\,{\rm dn}(\xi _{3})+
\frac{\chi _{s}(1-m_{3})^{2}+2\chi _{t}(1+m_{3})^{2}}{(1+m_{3})E(m_{3})-(1-m_{3})K(m_{3})}
\left( \frac{1}{1-m_{3}}\right) {\rm cn}(\xi _{3})\,{\rm dn}(\xi _{3})\times \nonumber \\
&&\Biggl(\left( \xi _{3}-\frac{1+m_{3}}{1-m_{3}}E({\rm am}(\xi
_{3}),m_{3})\right) +
\frac{{\rm sn}(\xi _{3})\,({\rm cn}^{2}(\xi _{3})+m_{3}^{2}{\rm dn}^{2}(\xi _{3}))}{\left(
1-m_{3}\right) {\rm cn}(\xi _{3})\,{\rm dn}(\xi _{3})}\Biggr)+  \nonumber \\
&&+2\chi _{t}\frac{1+m_{3}}{(1-m_{3})^{2}}\biggl((1+m_{3}){\rm cn}(\xi _{3})\,{\rm dn}(\xi
_{3})- (1+m_{3})+2m_{3}{\rm sn}^{2}(\xi _{3})\biggr).  \eqnum{38}
\end{eqnarray}] Here ${\rm cn}(\xi _{3})$ and ${\rm dn}(\xi _{3})$ are the elliptic
cosine and amplitude delta functions respectively. Their forms, properties
and the relations between them are given in \cite{16,17,18}, \cite{21,22}.
The details of the derivation of Eq.(38) are given in Appendix 2. It is seen
from Eq.(38) that the behaviour of $\chi _{zz}(\xi _{3})$ depends on $m_{3}$%
, which is the polarization ratio (see Eq. (19)). This dependence is
depicted in Fig.7a,b,c for two values of the susceptibility at the
interfaces: $\chi _{s}=2\chi _{t}$ and $\chi _{s}=0.$ The most interesting
general feature of $\chi _{zz}(\xi _{3})$ is the appearance of peaks as $%
m_{3}$ increases. The peaks become sharper and their maxima tend towards the
interfaces as $m_{3}$ approaches unity, which is when the maximum
polarization of the layer equals the thick film polarization limit. In this
limit $\chi _{zz}(\xi _{3})$ tends to $\chi _{t}$ in the major portion of
the film independent of the $\chi _{s}$ value, as it should in a thick film (%
$\alpha _{33}\rightarrow 0$, $P_{Azm}\rightarrow P_{Az}$). From this we see
that the solution (38) gives the correct value in the thick film limit.

For the electric field applied perpendicular to the multilayer surface, the
susceptibility $\chi _{zz}$ of the entire structure can be expressed as
\begin{equation}
\frac{1}{\chi _{zz}}=\frac{1}{L_{z}}\sum\limits_{j=0}^{N-1}\int%
\limits_{-L_{z}/2+2jl_{z}}^{-L_{z}/2+(2j+1)l_{z}}\frac{dz_{3}}{\chi
_{zz}(z_{3})}  \eqnum{39a}  \label{39a}
\end{equation}
This result follows from the fact that the capacitances $C_{i}$ of serially
connected capacitors obey the relation $1/C=\sum\limits_{i}1/C_{i}$. When
the electric field is applied parallel to x (which will interact with $%
P_{Bx} $ only) we have the result
\begin{equation}
\chi _{xx}=\frac{1}{L_{x}}\sum\limits_{j=0}^{N-1}\int%
\limits_{-L_{x}/2+(2j+1)l_{x}}^{-L_{x}/2+2(j+1)l_{x}}\chi _{xx}(z_{1})dz_{1}
\eqnum{39b}  \label{39b}
\end{equation}
since here the layers are connected in parallel. The inhomogeneous
susceptibility $\chi _{xx}(\xi _{1})$ of the B layers can, of course, be
obtained from Eq.(38) by substituting $z_{1}$ for $z_{3}$, $b_{1}$ for $%
a_{3} $, $\beta _{44}$ for $\alpha _{33}$ and $\xi _{1}$ for $\xi _{3}$. The
substitution of the dimensionless parameters $\xi _{3}$ for $z_{3}$ and $\xi
_{1}$ for $z_{1}$ in Eq.(39a) and Eq.(39b) respectively gives
\begin{equation}
\frac{1}{\chi _{zz}}=\frac{1}{2K(k_{Az})}\int\limits_{0}^{2K(k_{Az})}\frac{%
d\xi _{3}}{\chi _{zz}(\xi _{3})}  \eqnum{39c}  \label{39c}
\end{equation}

\begin{equation}
\chi _{xx}=\frac{1}{2K(k_{Bx})}\int\limits_{0}^{2K(k_{Bx})}\chi _{xx}(\xi
_{1})d\xi _{1}  \eqnum{39d}  \label{39d}
\end{equation}
Since the available experimental susceptibility data for multilayers \cite{5}
corresponds to the case of the electric field applied along the x direction,
we performed calculations on the basis of Eq.(39d). Details of these
calculations are given in Appendix 3. They yield the following expression
for the multilayer dielectric susceptibility $\chi _{xx}$:
\vspace*{10mm}
\begin{figure}[th]
\vspace*{5mm}
\centerline{\centerline{\psfig{figure=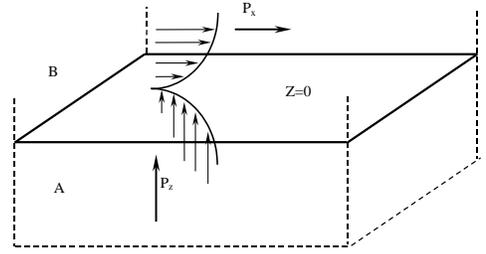,width=0.8\columnwidth}}}
\vspace*{15mm}
\caption{Variation of the polarization in a unit cell of a superlattice
with an ''a/c'' domain structure.}
\end{figure}
\vspace*{20mm}
\begin{figure}[th]
\centerline{\centerline{\psfig{figure=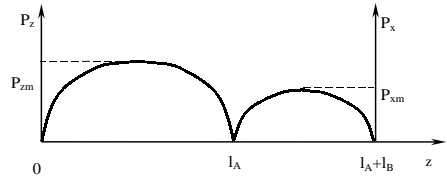,width=0.8\columnwidth}}}
\vspace*{10mm}
\caption{Periodic inhomogeneous polarization in a multilayer structure with
''a/c'' domains.}
\end{figure}

\twocolumn[\hsize\textwidth\columnwidth\hsize\csname @twocolumnfalse\endcsname
\begin{equation}
\chi _{xx}=\frac{1}{K(m_{1})}\frac{1+m_{1}}{(1-m_{1})^{2}}
\Biggl(\frac{(1-m_{1})^{2}\chi _{S}+2\chi _{t}(1+m_{1})^{2}}{(1+m_{1})E(m_{1})-(1-m_{1})K(m_{1})}
+2\chi _{t}((1-m_{1})K(m_{1})-2E(m_{1}))\Biggr), \eqnum{40}  \label{40}
\end{equation}] where $\chi _{t}^{-1}=4\left| b_{1}\right| $, $%
m_{1}=k_{Bx}^{2}$. Eq.(40) gives the dependence of the susceptibility on the
temperature, the polarization ratio, and the layer thickness via the
relation of these quantities (see Fig.4). In the thick film limit as $%
m_{1}\rightarrow 1,$ Eq.(40) gives $\chi _{xx}\rightarrow \chi _{t}$ (see
Appendix 4), which can also easily be obtained from Eq.(2b) (by neglecting
the gradient terms) as $\chi _{xx}^{-1}=d^{2}\Phi _{B}/dP_{Bx}^{2}$. The
susceptibility $\chi _{xx}$ corresponds to the ferroelectric phase of the
thin film multilayer, i.e. $T\leq T_{cl}^{B},$ where $T_{cl}^{B}$ is the
thickness induced ferroelectric phase transition temperature given by
Eq.(30). At this temperature the polarization is zero at the second order
phase transition, so that in the limit of $m_{1}\rightarrow 0$ we obtain
(see Appendix 4)
\begin{equation}
\chi _{xx}\rightarrow (\chi _{S}+2\chi _{t})8/3\pi ^{2}m_{1},\text{ }%
T\rightarrow T_{cl}^{B}  \eqnum{41}  \label{41}
\end{equation}
Hence we arrive at the divergence of the dielectric susceptibility for $%
T=T_{cl}^{B}$, which corresponds to $l_{Bx}=\pi .$ This same result can be
obtained directly by solving Eq.(37a) or its analog for $\chi _{xx}$ when $%
P_{Az}=0$ or $P_{Bx}=0.$ \ In particular, and keeping in mind that in the
considered layer structure any function should be periodic with the period
of the structure $2l$ (here we simplify things slightly by letting $%
l_{A}=l_{B}=l$ \cite{5,6,8}), we obtain:
\begin{equation}
\chi _{zz}(z_{3})=\chi _{0}+\frac{\sin (z_{3}+L_{z}/2-2jl_{Az})(\chi
_{s}-\chi _{0})}{\sin ({L_{z}/2-2jl_{Az})}}  \eqnum{42}  \label{42}
\end{equation}
where $\chi _{0}^{-1}=-2a_{0}^{A}(T_{cz}^{A}-T_{cl}^{A}).$

At the transition temperature $T=T_{cl}^{A}$ or $T_{cl}^{B}$ where $%
l_{z,x}=\pi $ and $L_{z,x}=2Nl_{z,x},$ one can see from Eq.(42) that $\chi
_{zz}(z)$ and $\chi _{xx}(z)$ become infinitely large because $\sin (N\pi
-2\pi j)=0$. The divergence of the susceptibility at the temperature of the
thickness induced phase transition is similary obtained for $\chi _{zz}$\
and $\chi _{xx}$\ after integration and summation of Eqs.(39a) and (39b).
Therefore the divergence of the susceptibility $\chi _{zz}$ or $\chi _{xx}$
at the transition temperature $T=T_{cl}^{A}$ or $T=T_{cl}^{B}$ is the {\it %
characteristic} feature of ferroelectric thin film multilayers. In Fig.8a we
display the temperature dependence of the susceptibility in the
ferroelectric region $T<T_{cl}^{B}$ for the thickness $l_{B}=2l_{B0},$
(recall that $l_{B0}$\ is the thickness where $T_{cl}^{B}$\ vanishes) which
corresponds to $T_{cl}^{B}/T_{cx}^{B}=0.75$ (see Eq.(30)) and $\chi _{s}=0.$
Since $l_{B}=l_{B0}$ gives $T_{cl}^{B}=0$ this ratio will be smaller than
0.75 at $l_{B0}<l_{B}<2l_{B0}$ and larger than 0.75 at $l_{B}>2l_{B0}.$ It
appears possible to write the temperature dependence given by Eq.(40) in the
form of a Curie-Weiss (C-W) law at $T\rightarrow
T_{cl}^{B}\;(m_{1}\rightarrow 0)$. We obtain from Eq.(41) the following
approximate forms of $\chi _{xx}(T)$for three different values of $\chi _{s}$%
:

\begin{equation}
\chi _{xx}(T)\simeq \frac 2{\pi ^2a_0^B}\frac 1{T_{cl}^B-T}=\frac{c_B}{%
T_{cl}^B-T};\text{ \ \ \ \ }\chi _s=0  \eqnum{43}
\end{equation}

\begin{eqnarray}
&&\chi _{xx}(T)\approx \left( 4\chi _{s}\frac{\beta _{44}}{l_{B}^{2}}+\frac{2%
}{\pi ^{2}}\right) \frac{1}{a_{0}^{B}\left( T_{cl}^{B}-T\right) };  \nonumber
\\
&&\chi _{s}=const\neq 0  \eqnum{44}
\end{eqnarray}
\begin{eqnarray}
&&\chi _{xx}(T)\approx \frac{2+\alpha }{\pi ^{2}}\frac{1}{a_{0}^{B}\left(
T_{cl}^{B}-T\right) };  \nonumber \\
&&\chi _{s}=\alpha \chi _{t}=\frac{\alpha }{4a_{0}^{B}\left(
T_{cx}^{B}-T\right) }  \eqnum{45}
\end{eqnarray}
Details of these calculations are given in Appendix 5. Note that the C-W
constant for $\chi _{s}=0$\ is very close to that of the thick film value, $%
C_{t}=(4a_{0}^{B})^{-1}.$\ It is seen from Fig.8a that Eq.(43) fits
surprisingly well the temperature dependence given by Eq.(40) (see solid
curve in Fig.8a) not only in the vicinity of $T=T_{cl}^{B}$\ but also for
the entire temperature region. Eq.(45) also gives a good fit after
renormalization of the $\chi _{1}$\ value in Fig.8a. Eqs.(43) and (45) show
that the susceptibility dependence on the film thickness is mainly defined
by the thickness dependence of $T_{cl}^{B}.$\ We point out that for $\chi
_{S}=const\neq 0,$ $C_{B}$ is also thickness dependent (see Eq.(44)).

Well into the paraelectric phase at $T>T_{cl}^{A}$ or $T>T_{cl}^{B}$ where
there is zero spontaneous polarization, there will be some small homogeneous
polarization induced by the external electric field. In this case the free
energy can be expanded in a power series of the polarization, and the static
dielectric susceptibility can be found, as usual, by:
\begin{equation}
\chi _{zz}^{-1}=\frac{\partial ^{2}\Phi }{\partial P_{z}^{2}};%
\text{ \ \ \ }\chi _{xx}^{-1}=%
\frac{\partial ^{2}\Phi }{\partial P_{x}^{2}}  \eqnum{46}
\end{equation}
Eq.(46) leads to conventional C-W laws for the susceptibilities in the
paraelectric phase of thin films which supplement the susceptibility in the
ferroelectric phases of the ''A'' and ''B'' layers
\begin{eqnarray}
\chi _{zz} &=&\frac{1}{c_{0}^{A}(T-T_{cl}^{A})};\text{ \quad }T\geq
T_{cl}^{A}  \eqnum{47} \\
\chi _{xx} &=&\frac{1}{c_{0}^{B}(T-T_{cl}^{B})};\text{ \quad }T\geq
T_{cl}^{B}  \eqnum{48}
\end{eqnarray}
{\it \ }which diverge at the transition temperatures $T_{cl}^{A}$ or $%
T_{cl}^{B}$, which can be considered as the Curie temperatures of the
thickness induced ferroelectric phase transition with the characteristic
dependence on the film thickness (see Fig.5). Note that for thick films,
when $T_{cl}^{A}\rightarrow T_{c0}^{A}$ and $T_{cl}^{B}\rightarrow
T_{c0}^{B} $ (see Fig.5), the coefficients $c_{0}^{A}\rightarrow a_{0}^{A}$
and{\it \ }$c_{0}^{B}\rightarrow a_{0}^{B}$, but in general $c_{0}$ and $%
a_{0}$ are different. Unfortunately it is cumbersome to compare the values
of $c_{B}$ \ in Eq.(43) with those of $c_{0}^{B}$ in Eq.(48). For the sake
of illustration we depict in Fig.8b the $\chi _{xx}$ dependence for the
entire temperature region in supposing that $c_{B}=1/(2c_{0}^{B})${\it \ }and%
{\it \ }$c_{B}=2/c_{0}^{B}$. The divergence at $T=$ $T_{cl}^{B}$ is the most
significant feature here.

It will be shown later, by comparison with the experimental data \cite{5},
that the giant dielectric response observed in some multilayers originates
from the susceptibility anomaly related to thickness induced phase
transition considered above. Of course the internal fields induced by misfit
dislocations, growth imperfections and impurities will smear out the
dielectric response of the multilayers and thus reduce infinity to a more
realistic value.

We will now discuss the possibility of calculating the thin film
susceptibility as the second derivative of the free energy instead of
solving the Lam\'{e} equation. To do this we assume that in the
ferroelectric phase the role of the second term in Eq.(37) will increase
with increasing polarization, and so it should be possible to neglect the
contribution of the second derivative. In this case the susceptibility
becomes $\chi _{zz}^{-1}\approx 2a_{3}+12a_{33}P_{Az}^{2}$. Integration of
this expression over the layer thickness (with $P_{Az}(z)$ given by Eq.(23))
and then summing over the layers leads to the following expression
\begin{equation}
\frac{1}{\chi _{zz}}=2a_{3}\left[ {1-\frac{6}{1+k_{Az}^{2}}\left( {1-\frac{%
E(k_{Az})}{K(k_{Az})}}\right) }\right]  \eqnum{49}  \label{49}
\end{equation}
where $k_{Az}$ is given by Eq.(19).

Equation (49) gives the correct expression for the susceptibility in the
thick film paraelectric phase ($k_{Az}=0$, $\chi _{zz}^{-1}=2a_{3}$, $%
T>T_{c0}^{A}$) and in the thick film ferroelectric phase ($k_{Az}=1,$ $\chi
_{zz}^{-1}=-4a_{3}$, $T<T_{c0}^{A}$). However, problems arise with this
approximation for intermediate values of $k_{Az}.$ For $%
k_{Az}<k_{Az}^{0}=0.674$ we obtain a negative expression for $\chi _{zz}$
(including the region of thin film paraelectric phase $T\geq T_{cl}^{A}$)
and a positive expression when $1\geq k_{Az}>k_{Az}^{0}$ with $\chi _{zz}$
diverging at $k_{Az}^{0}=0.674$ ($k_{Az}^{0}$ sends the expression (49) for $%
\chi _{zz}^{-1}$ to zero). It is obvious that for $k_{Az}<k_{Az}^{0}$ the
susceptibility must be calculated from the differential equation Eq.(37)
instead of the approximation derived Eq.(49) since the latter leads to the
physically unreasonable negative susceptibility. The validity of Eq.(49) in
the region $k_{Az}^{0}<k_{Az}\leq 1$ when $1/\chi _{zz}>0$ can be checked by
a $1/\chi _{zz}$ calculation on the basis of the exact solution of Eq.(37).
Thus we see that Eq.(49) will be valid only in the thick film limit of $%
k_{Az}=1$. Therefore, for the majority of thin films and multilayers, the
dielectric susceptibility will be defined by Eq.(37) whereas the normally
used free energy second derivative (see, e.g. \cite{9}) is valid in the
ferroelectric phase only for thick films. We also point out that Eq.(37)
will be applicable to bulk ferroelectrics with inhomogeneous polarization,
i.e. when there is a strong contribution of the polarization gradient since
the mean field approximation is no longer valid.

\section{Dynamic properties}

\subsection{Polarization}

To consider the dynamic properties of a periodic spatial structure, we
should add the time derivatives of the polarization to equations (12a) and
(12b). This can be done by the procedure suggested in \cite{22,23}. This
procedure permits the investigation of low frequency dynamics in
ferroelectrics. Its essence is to expand the polarization in powers of
frequency or the time derivative operator $d/dt$. The first order term
gives, as usual, the decay of the polarization, while the second order term
provides an ''oscillatory response'' and contains the mass coefficient $\mu $%
.

The equations of motion can be written as

\begin{equation}
\mu _{A}\frac{\partial ^{2}P_{Az}}{\partial t^{2}}+%
\gamma _{A}\frac{\partial P_{Az}}{\partial t}+\frac{%
\delta F_{A}}{\delta P_{Az}}=0  \eqnum{50a}  \label{50a}
\end{equation}

\begin{equation}
\mu _{B}\frac{\partial ^{2}P_{Bx}}{\partial t^{2}}+%
\gamma _{B}\frac{\partial P_{Bx}}{\partial t}+\frac{%
\delta F_{B}}{\delta P_{Bx}}=0  \eqnum{50b}  \label{50b}
\end{equation}

The free energy variation $\delta F/\delta P$ in the Euler-Lagrange
Eqs.(12a) and (12b), yields the equations for $P_{Az}$ and $P_{Bx}$
\vspace*{-3mm}
\begin{figure}[th]
\centerline{\centerline{\psfig{figure=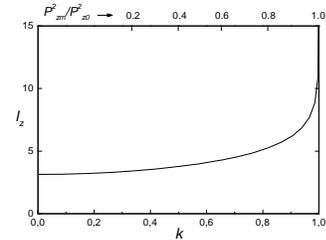,width=0.7\columnwidth}}}
\caption{Dimensionless thickness of an ''A'' or a ''B'' layer versus the
polarization ratio.}
\end{figure}
\vspace*{5mm}
\begin{figure}[th]
\centerline{\centerline{\psfig{figure=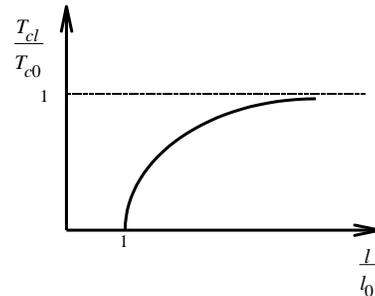,width=0.7\columnwidth}}}
\vspace*{-8mm}
\caption{Temperature of the thickness induced ferroelectric phase
transition as a function of layer thickness.}
\end{figure}

\twocolumn[\hsize\textwidth\columnwidth\hsize\csname @twocolumnfalse\endcsname
\begin{equation}
\mu _{A}\frac{\partial ^{2}P_{Az}}{\partial t^{2}}+\gamma _{B}\frac{\partial
P_{Az}}{\partial t}+a_{3}P_{Az}+2a_{33}P_{Az}^{3}-\alpha _{33}\frac{\partial
^{2}P_{Az}}{\partial z^{2}}=0  \eqnum{51a}  \label{51a}
\end{equation}

\begin{equation}
\mu _{B}\frac{\partial ^{2}P_{Bx}}{\partial t^{2}}+\gamma _{B}\frac{\partial
P_{Bx}}{\partial t}+b_{1}P_{Bx}+2b_{11}P_{Bx}^{3}-\beta _{44}\frac{\partial
^{2}P_{Bx}}{\partial z^{2}}=0  \eqnum{51b}  \label{51b}
\end{equation}]

Since Eqs. (51a) and (51b) are essentially the same we will consider the
solution for $P_{Az}$ keeping in mind that $P_{Bx}$ can be written, as
usual, by substitution of the relevant coefficients and constants in Eq.(51).

We will look for solutions of Eq.(51a) in the form of stationary nonlinear
waves in the conventional self-similar form

\begin{equation}
P_{Az}=P_{Az}(z-vt)=P_{z}(kz-\omega t),\quad \text{ }\omega =kv  \eqnum{52}
\label{52}
\end{equation}
where $\omega $, $k$ and $v$ are the frequency, the wave vector, and the
velocity respectively and both vectors are directed along the $z$ axis.
Since the wave propagates along the normal to the multilayer structure ($k$
and $v$ are parallel to $z$) we have suppressed the index $z$ in $k$ and $v$
in Eq.(52). Using the relations

\begin{equation}
\xi =z-vt;\;\frac{\partial }{\partial t}=-v\frac{d}{d\xi };\text{ }\frac{%
\partial ^{2}}{\partial t^{2}}=v^{2}\frac{d^{2}}{d\xi ^{2}};\text{ }\frac{d^{2}}{dz^{2}}=%
\frac{d^{2}}{d\xi ^{2}}  \eqnum{53}  \label{53}
\end{equation}
we obtain from (51a)

\begin{equation}
(\mu _{A}v^{2}-\alpha _{33})\frac{d^{2}P_{Az}}{d\xi ^{2}}-\gamma _{A}v\frac{%
dP_{Az}}{d\xi }+a_{3}P_{Az}+2a_{33}P_{Az}^{3}=0  \eqnum{54}  \label{54}
\end{equation}

Since we are interested in the dispersion law for nonlinear waves (which at $%
t=0$ gives us our static periodic structure (23)) we shall neglect the decay
term. In doing so ($\gamma _{A}=0$) Eq.(54) coincides with Eq.(12a) but with
$\alpha _{33}\rightarrow \alpha _{33}-\mu _{A}v^{2}$, so that the solution
of Eq.(54) can be written in the form of Eq.(23) :

\begin{equation}
P_{Az}(z-vt)=P_{Azm}{\rm sn}\left( \frac{\overline{z_{3}}-\overline{v}t+\overline{%
L_{z}}/2-2j\overline{l_{Az}}}{\sqrt{1+k_{Az}^{2}}},k_{Az}\right)  \eqnum{55}
\label{55}
\end{equation}
where $\overline{v}=\sqrt{-a_{3}/(\alpha _{33}-\mu _{A}v^{2})}v$ and the
renormalized values of $\overline{z_{3}}$, $\overline{L_{z}}$, and $%
\overline{l_{Az}}$ can easily be obtained from (22) by substituting $\sqrt{%
-a_{3}/(\alpha _{33}-\mu _{A}v^{2})}$ for $\sqrt{-a_{3}/\alpha _{33}}$.

The solution of Eq.(51b) can be similarly obtained and written in the form:

\begin{equation}
P_{Bx}(z-vt)=P_{Bxm}{\rm sn}\left( \frac{\overline{z_{1}}-\overline{v}t+\overline{%
L_{x}}/2-2j\overline{l_{Bx}}}{\sqrt{1+k_{Bx}^{2}}},k_{Bx}\right)  \eqnum{56}
\label{56}
\end{equation}
with $\overline{v}=\sqrt{-b_{1}/(\beta _{44}-\mu _{B}v^{2})}v$ and $%
\overline{z_{1}}$, $\overline{L_{z}}$, $\overline{l_{Az}}$ follow from
Eq.(29) after \ the substitution of $\sqrt{-b_{1}/(\beta _{44}-\mu _{B}v^{2})%
}$ for $\sqrt{-b_{1}/\beta _{44}}$. Eqs. (55) and (56) give the
time-dependent inhomogeneous polarization of a multilayer structure.

\subsection{Dispersion law}

To find the dispersion law of the stationary nonlinear waves formed from
Eqs. (55) and (56) in the A and B layers respectively we first recall that
the wave number $k=2\pi /(\lambda _{Az}+\lambda _{Bx})$, where $\lambda
_{Az} $ and $\lambda _{Bx}$ are the wavelengths of the $P_{Az}$ and $P_{Bx}$
nonlinear waves respectively. They are actually determined by Eqs.(25) and
(28b) with respect to the substitution $\alpha _{33}(or\,\beta
_{44})\rightarrow \alpha _{33}(or\,\beta _{44})-\mu _{A,B}v_{A,B}^{2}$. Hence

\begin{eqnarray}
\lambda _{Az} &=&2\sqrt{\frac{\alpha _{33}-\mu _{A}v^{2}}{-a_{3}}}\sqrt{%
1+k_{Az}^{2}}K(k_{Az})  \nonumber \\
\lambda _{Bx} &=&2\sqrt{\frac{\beta _{44}-\mu _{B}v^{2}}{-b_{1}}}\sqrt{%
1+k_{Bx}^{2}}K(k_{Bx})  \eqnum{57}  \label{57}
\end{eqnarray}

It then follows that
\vspace*{10mm}
\begin{figure}[th]
\vspace*{10mm}
\centerline{\centerline{\psfig{figure=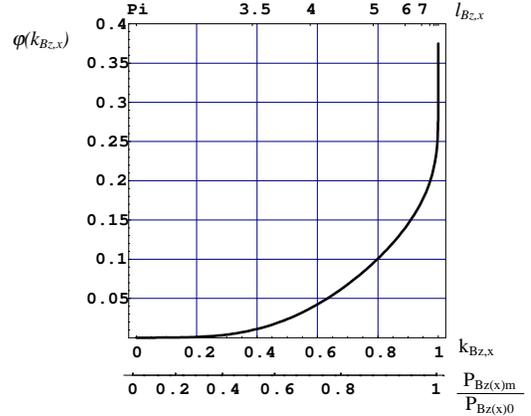,width=0.8\columnwidth}}}
\vspace*{8mm}
\caption{Form of the function $\protect\varphi (k)$ in Eq.(32) which
determines the criterion for the multilayer domain structure. }
\end{figure}

\twocolumn[\hsize\textwidth\columnwidth\hsize\csname @twocolumnfalse\endcsname
\begin{equation}
k=\frac{2\pi }{\lambda _{Az}+\lambda _{Bx}}=\frac{\pi }{\sqrt{(\alpha
_{33}-\mu _{A}v^{2})/(-a_{3})}\sqrt{1+k_{Az}^{2}}K(k_{Az})+\sqrt{(\beta
_{44}-\mu _{B}v^{2})/(-b_{1})}\sqrt{1+k_{Bx}^{2}}K(k_{Bx})}  \eqnum{58}
\label{58}
\end{equation}] Expression (58) implicitly determines the dispersion law,
since $v=\omega /k$, and so Eq.(58) can give the $\omega (k)$ dependence.
Let us simplify Eq. (58). It follows from Eqs. (25) and (28b) that

\begin{eqnarray}
&&l_{A}=2\sqrt{\frac{\alpha _{33}}{-a_{3}}}\sqrt{1+k_{Az}^{2}}K(k_{Az});
\nonumber \\
&&l_{B}=2\sqrt{\frac{\beta _{44}}{-b_{1}}}\sqrt{1+k_{Bx}^{2}}K(k_{Bx})
\eqnum{59}  \label{59}
\end{eqnarray}
Thus we can rewrite Eq.(58) as:

\begin{equation}
k=2\pi \frac{1}{l_{A}\sqrt{1-\mu _{A}v^{2}/\alpha _{33}}+l_{B}\sqrt{1-\mu
_{B}v^{2}/\beta _{44}}}  \eqnum{60a}  \label{60a}
\end{equation}
or

\begin{equation}
l_{A}\sqrt{k^{2}-\frac{\mu _{A}}{\alpha _{33}}\omega ^{2}}+l_{B}\sqrt{k^{2}-%
\frac{\mu _{B}}{\beta _{44}}\omega ^{2}}=2\pi  \eqnum{60b}  \label{60b}
\end{equation}

To facilitate the analysis of the dispersion law given by Eqs.(60) we will
first assume that $\eta =\eta _{A}=\eta _{B}$, where $\mu _{A}/\alpha
_{33}=\eta _{A}$, $\mu _{B}/\beta _{44}=\eta _{B}$ and $l_{A}=l_{B}=l$. In
this case we have from Eq.(60):

\begin{equation}
\omega ^2=\frac 1\eta \left[ k^2-\frac{\pi ^2}{l^2}\right]  \eqnum{61}
\label{61}
\end{equation}

It is seen from Eq.(61) that the dispersion law has the familiar long
wavelength form. It also exhibits the peculiar dependence of the nonlinear
wave frequency on its amplitude via relation between $l$ and $k_{z,x}$
(Eq.(59)), see e.g. \cite{25}.

We schematically plot the dispersion law in Fig.9a. First of all since $%
\omega ^{2}>0$ we have some critical value $k=k_{c}=\pm \pi /l$ for which $%
\omega =0$ .

On the other hand for any specific value $k=k_{0}$ there is a critical
thickness $l_{c}=\pm \pi /k_{0}$ at which $\omega =0$ and $\omega \neq 0$
for $l>l_{c}$ only. The thickness dependence of the frequency has the form:

\begin{equation}
\omega =\pm \sqrt{\frac{1}{\eta }}\frac{k_{0}}{l}\sqrt{l^{2}-l_{c}^{2}}
\eqnum{62}  \label{62}
\end{equation}
which in the vicinity of $l=l_{c}$ gives $\omega \approx \pm \sqrt{1/\eta }%
k_{0}\sqrt{2/l_{c}}\sqrt{l-l_{c}}$. Thus the frequency increases as the
square root of\ $l$ . This dependence is represented in Fig.9b.

For the general anisotropic case, $\eta _{A}\neq \eta _{B}$ and $\ \omega
(l) $ and $\omega (k)$ will be represented by more complicated expressions.
They can be derived by squaring both sides of Eq.(60b). We obtain for $%
l_{A}=l_{B}=l$:

\begin{figure}[th]
\vspace*{10mm}
\centerline{\centerline{\psfig{figure=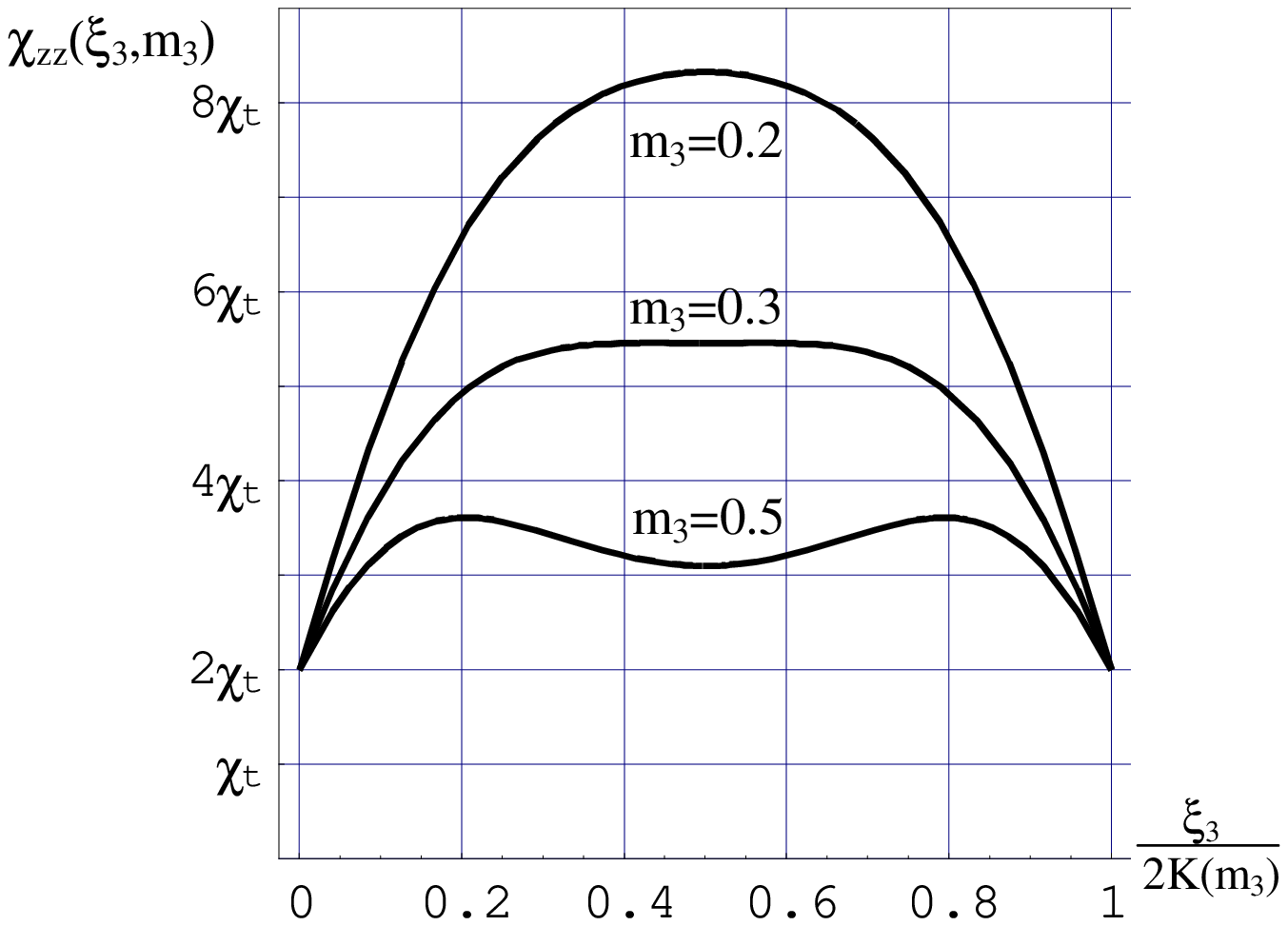,width=0.8\columnwidth}}}
 \center {\text{a}}
\centerline{\centerline{\psfig{figure=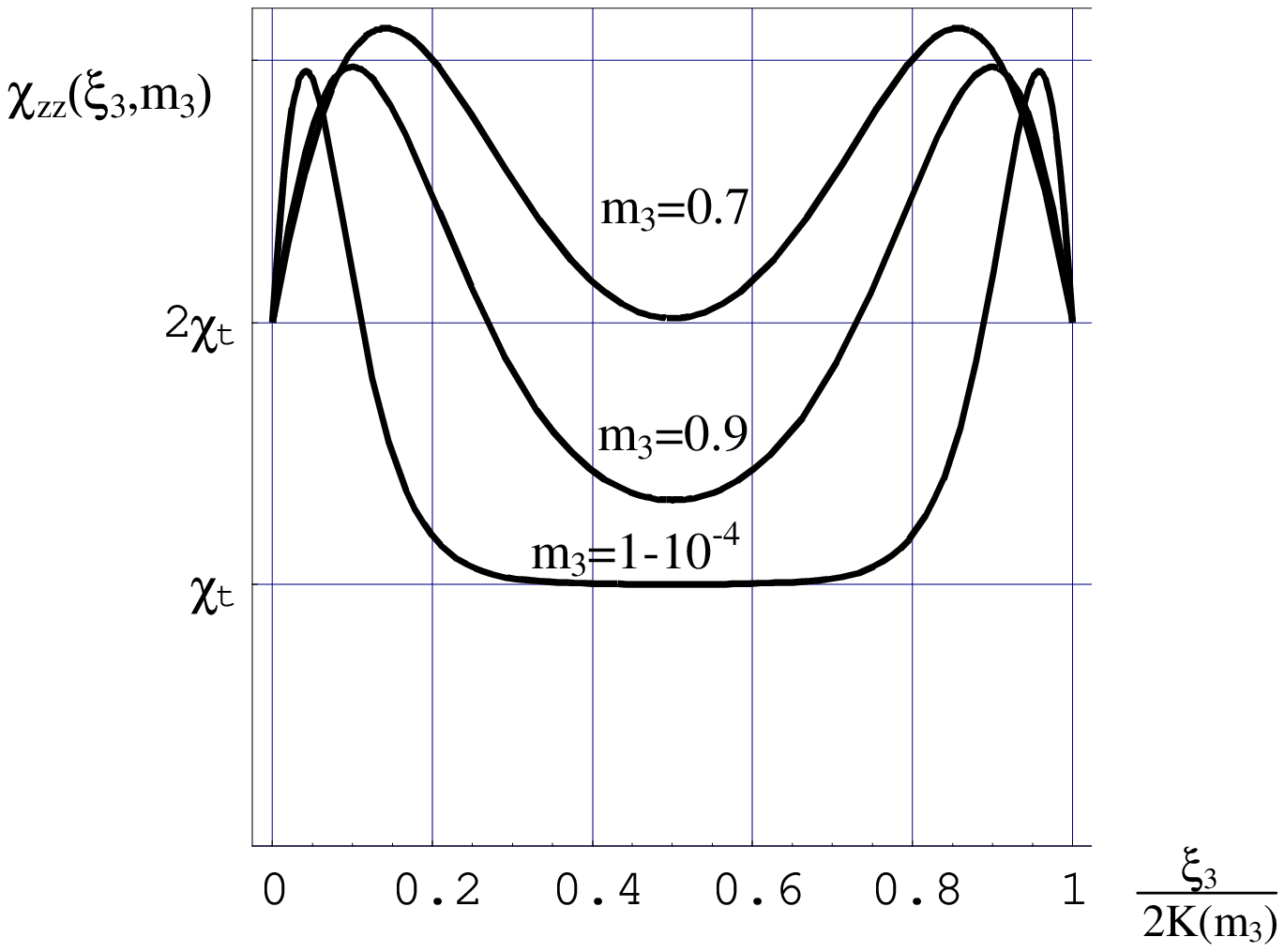,width=0.8\columnwidth}}}
\center { \text{b}}
\centerline{\centerline{\psfig{figure=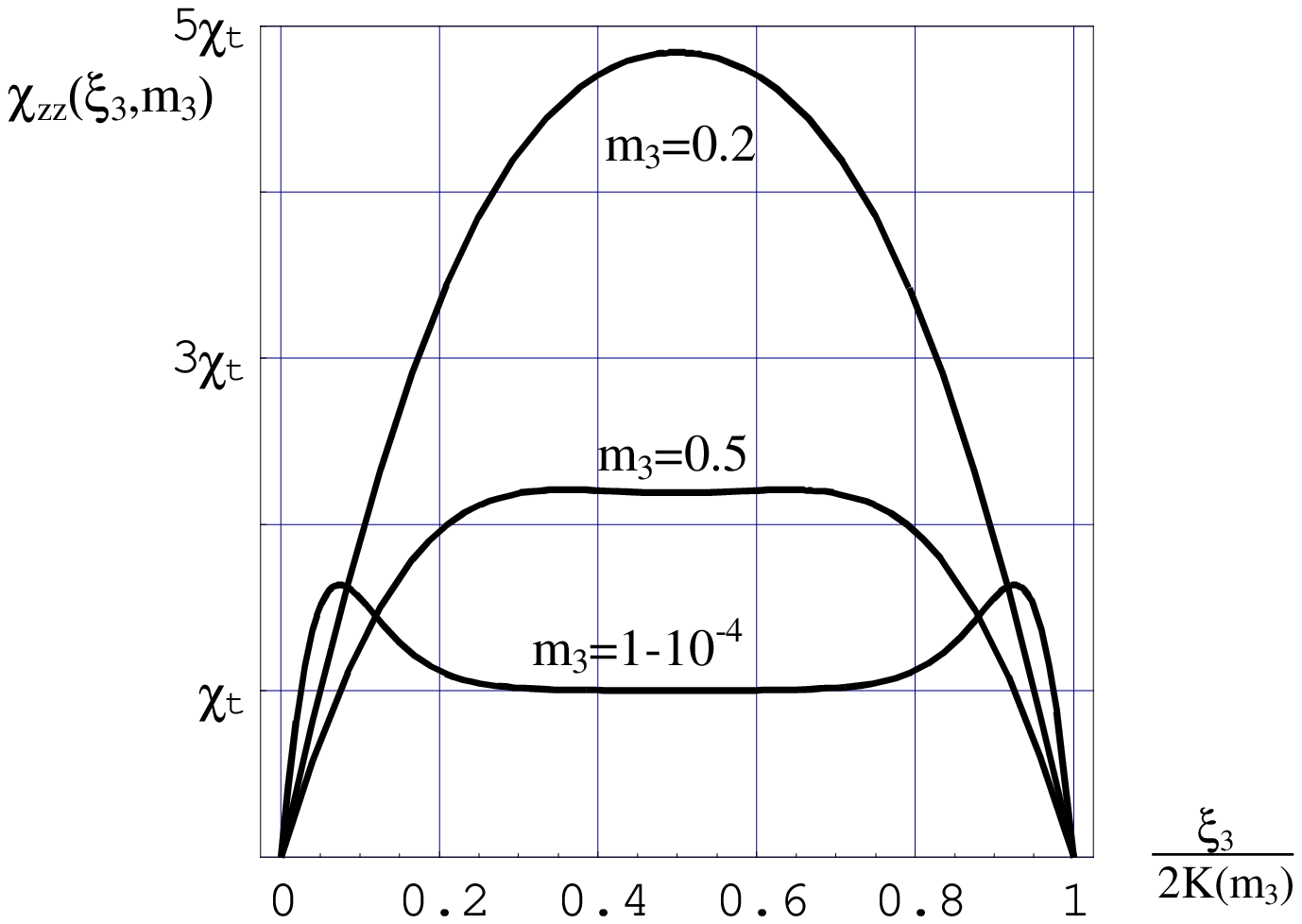,width=0.8\columnwidth}}}
\center {\text{c}}
\vspace*{8mm}
\caption{Inhomogeneous susceptibility dependence on the coordinate $
\xi _{3}/(2K(m_{3}))=z_{1}/l_{Az}$ for different $m_{3}$ values for $
\chi _{S}=2\chi _{t}$ (a and b) and for $\chi _{S}=0$ (c). }
\end{figure}

\twocolumn[\hsize\textwidth\columnwidth\hsize\csname @twocolumnfalse\endcsname
\begin{equation}
\omega _{1,2}^{2}=\frac{1}{2(\eta _{Az}-\eta _{Bx})^{2}}\left\{ -\frac{8\pi
^{2}}{l^{2}}(\eta _{Az}+\eta _{Bx})+\frac{4\pi }{l}\sqrt{\frac{4\pi ^{2}}{l^{2}}(\eta _{Az}+\eta _{Bx})^{2}-4\left( \frac{\pi ^{2}}{l^{2}}-k^{2}\right) (\eta _{Az}-\eta _{Bx})^{2}}\right\}  \eqnum{63}  \label{63}
\end{equation} ]

We have dropped the minus sign before the square root in Eq.(63) because $%
\omega _{1,2}^{2}>0$. It is seen that when $k^{2}=\pi ^{2}/l^{2}$ $\ $then$\
\omega _{1,2}=0$,\ similar to the isotropic case. Thus this noteworthy
aspect of the dispersion law is conserved for the general anisotropic case.
We plot the general dependence in Fig.10. Although the qualitative features
of the $\omega (k)$ law for the isotropic case are preserved in the
anisotropic case, there are peculiarities at large $k$ in the anisotropic
case: we determine $\omega \sim \sqrt{k}$ rather than linear dependence in
the isotropic case (compare Figs.10a and 9a) and there is a different
thickness dependence, with a small decrease of $\omega (l)$ at large $l$
(compare Figs. 10b and 9b).
\vspace*{5mm}
\begin{figure}[tbp]
\centerline{\centerline{\psfig{figure=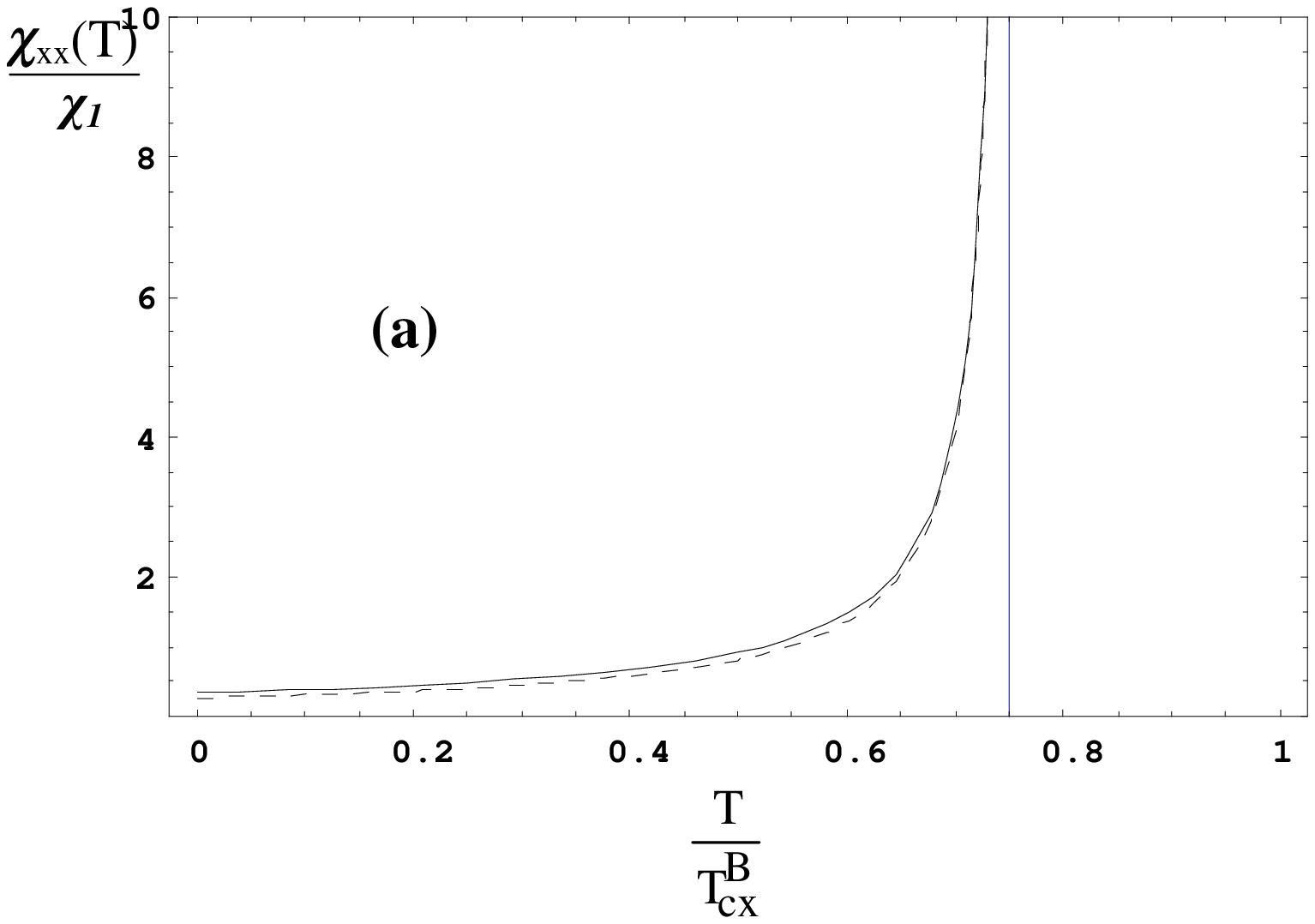,width=0.8\columnwidth}}}
\centerline{\centerline{\psfig{figure=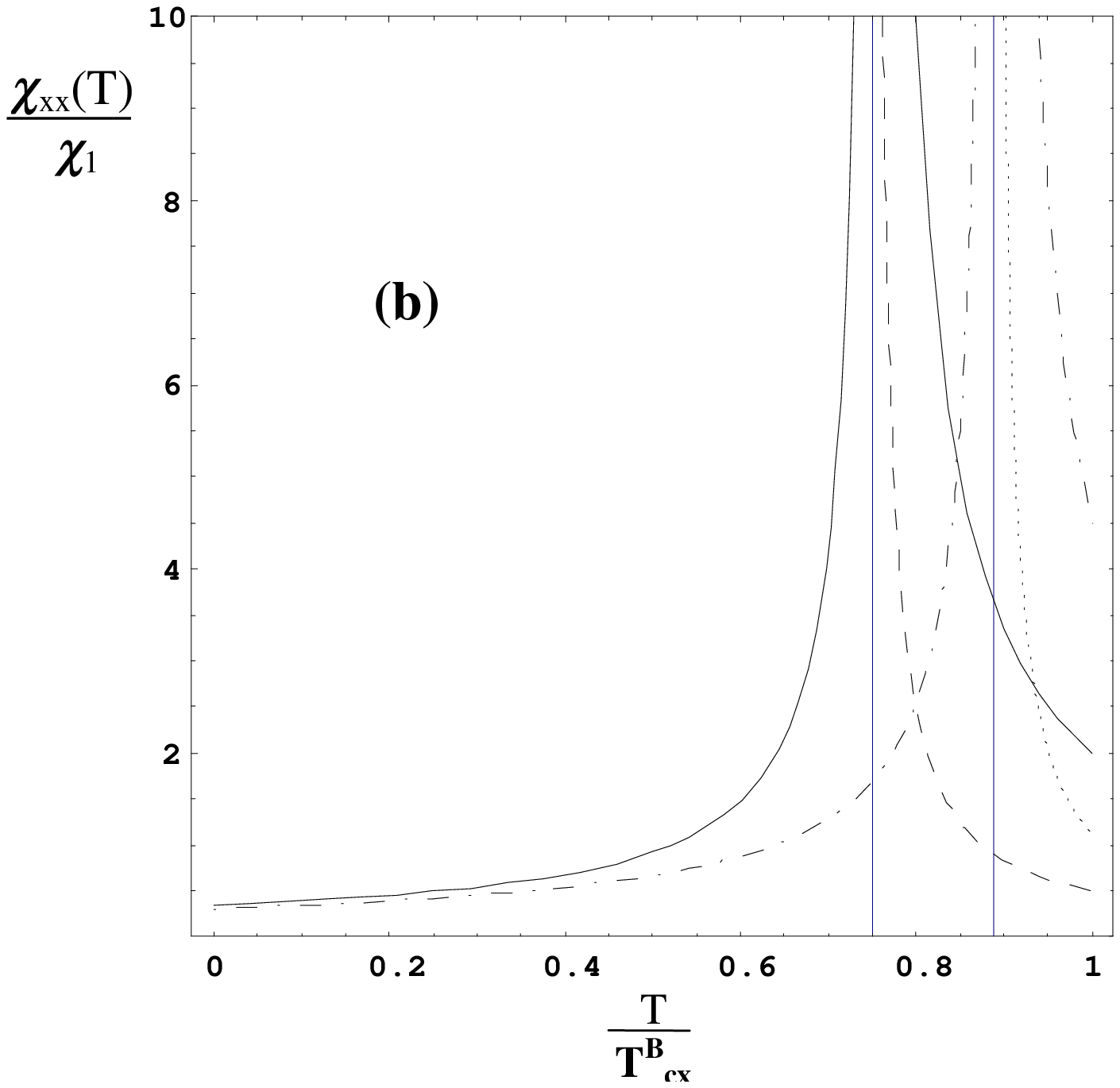,width=0.8\columnwidth}}}
\vspace*{5mm}
\caption{Temperature dependence of the dielectric susceptibility in units
of $\chi _{1}=1/(a_{0}^{B}T_{cx}^{B})$, for $\chi _{S}=0$
(a) at $T\leq T_{cl}^{B}$ using Eq.(40) (solid line) and Curie-Weiss law
(dashed line) for $l_{B}=2l_{B0}$ and (b) in the entire temperature range
using Eq.(40) ($T<T_{cl}^{B})$ and Eq.(48) ($T>T_{cl}^{B}$) for different $%
c_{0}^{B}$ values and for $l_{B}=2l_{B0}$ (solid line: $1/c_{0}^{B}=2c_{B}$;
dashed line: $1/c_{0}^{B}=c_{B}/2$) and $l_{B}=3l_{B0}$ (dot-dashed line : $%
1/c_{0}^{B}=2c_{B}$; dotted line: $1/c_{0}^{B}=c_{B}/2$).}
\end{figure}
\vspace*{3mm}
\begin{figure}[th]
\centerline{\centerline{\psfig{figure=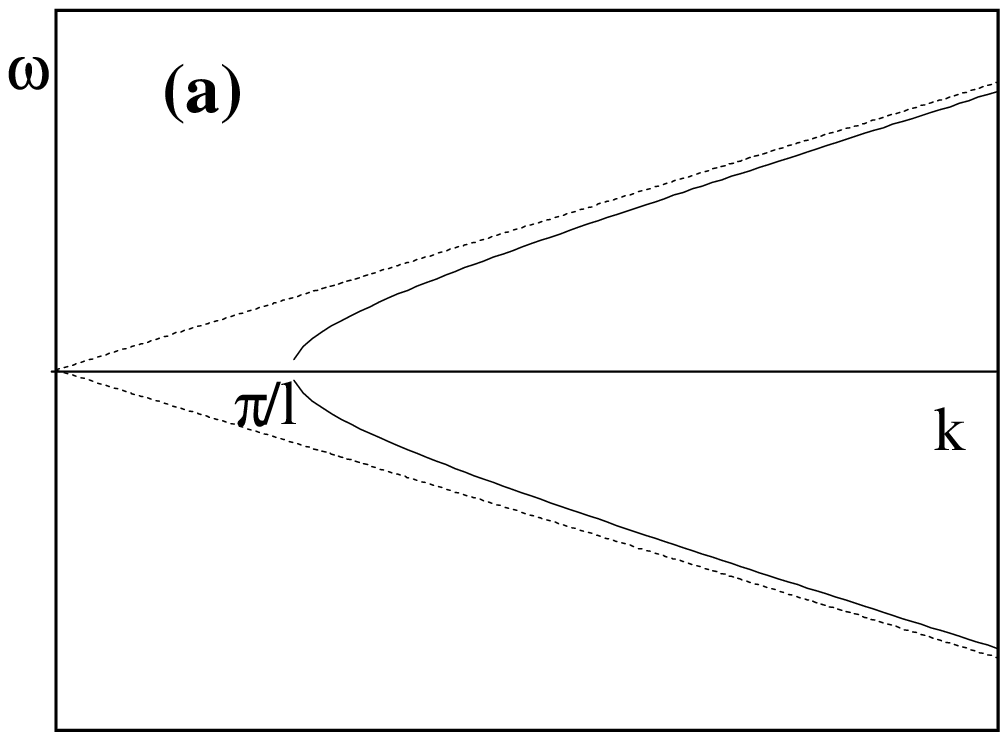,width=0.7\columnwidth}}}
\vspace*{7mm}
\centerline{\centerline{\psfig{figure=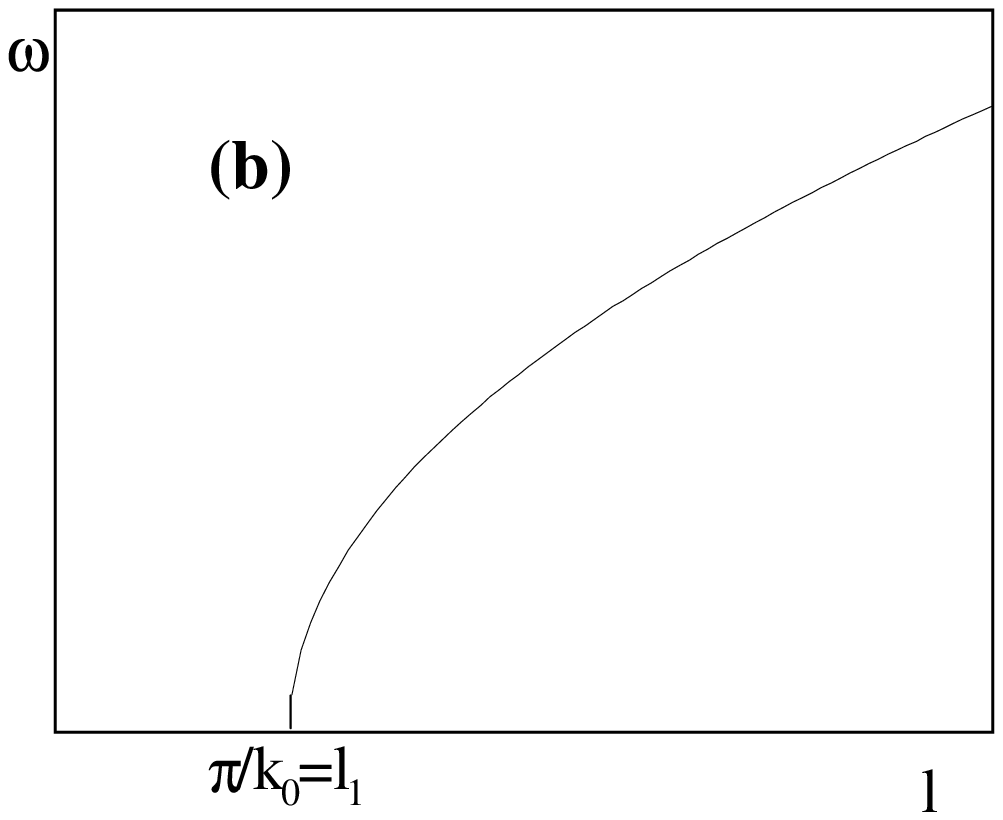,width=0.7\columnwidth}}}
\vspace*{5mm}
\caption{Nonlinear wave dispersion law for a multilayer in the isotropic
case as a function of (a) the wave vector and (b) the film thickness. The
dashed line is $\omega =k\eta ^{-1/2}$.}
\end{figure}

\section{Discussion. Giant dielectric response of ferroelectric thin film
multilayers.}

Let us begin with the comparison of the calculated and the observed
dielectric susceptibility. The susceptibility was recently measured \cite{5}
in superlattices consisting of ferroelectric PbTiO$_{3}$ and paraelectric Pb$%
_{1-x}$La$_{x}$TiO$_{3}$ ($x=0.28$ - PLT) grown on (100)-oriented SrTiO$_{3}$
single crystal substrates. Three superlattices of PT/PLT with modulation
wavelengths of 100 $A^{\circ }$ (sample S-40), 400 $A^{\circ }$ (sample
S-10), 2000 $A^{\circ }$ (sample S-2) were studied. In each superlattice the
PT and PLT layers were of equal thickness, and the total thickness was 4000 $%
A^{\circ }$. The authors observed a Debye-like frequency dispersion of the
real and imaginary parts of the dielectric susceptibility $\varepsilon _{xx}$
for the samples S-10 and S-2 in which the value of the real part of the
dielectric susceptibility at low frequency approached 420000 and 350000
respectively at $T\approx 50^{o}C$. A significant increase in the
susceptibility with temperature was observed in sample S-10. We present this
temperature dependence by the full circles in Fig.11, where the solid line
represents the C-W law given by Eq.(43) with $c_{B}=7\cdot 10^{6}K$, and $%
T_{cl}^{B}($S-10$)=533K$. The independence of $c_{B}$ on layer thickness
(see Eq.(43)) makes it possible to obtain\ the transition temperature for
specimen S-2: $T_{cl}^{B}($S-2$)=575K$. Taking the ratio of these $%
T_{cl}^{B} $ values and using Eq.(30) for S-10 and S-2 leads to the critical
thickness $l_{B0}=55A^{o}.$ This value is larger than the layer thickness ($%
l_{B}=50A^{o}$) in S-40 specimen, which is why no ferroelectric phase
transition was observed for this multilayer. Thus the value reported for
S-40, $\varepsilon _{xx}^{\prime }\simeq 750$\ at $T\approx 50^{o}C$,
represents the contribution from the paraelectric phases of both the PT and
PLT layers. This value is several orders of magnitude smaller than the
values measured in the ferroelectric phase for S-10 and S-2. The
contribution from the PT-layers in the paraelectric phase of the S-40
sample, where $T_{cl}\approx 0$\ (see Eq.(30) at $l_{B}\approx l_{B0}$) can
be written as $\varepsilon _{xx}=4\pi c_{Bp}/T$, where $c_{Bp}=c_{0}^{B}$\
in accordance with Eq.(48). Keeping in mind that $\varepsilon _{xx}^{\prime
}(PT)<750$, we can conclude that the C-W constant in Eq.(48) $c_{Bp}<2\cdot
10^{4}$. Therefore the dielectric susceptibility in the paraelectric phase
of the thin film is much smaller than in the ferroelectric phase. This is
contrary to what is observed in bulk ferroelectrics. On the other hand this
gives support to the supposition that the contribution of the permittivity
from the PLT paraelectric phase to the giant value observed in S-10 and S-2
is negligibly small in comparison with the contribution from the PT layers.
Estimation of the transition temperature for a thick layer, $T_{cx}^{B}$,
with the help of Eq.(30) and the values determined above for $T_{cl}^{B}$
leads to $T_{cx}^{B}=580K$. This value is smaller than that for bulk PbTiO$%
_{3}$ : $T_{c0}^{B}=763K$. This would seem to favor compressive strain in
the layer, i.e. $X^{B}<0$ in Eq.(9b) because $Q_{11}^{B}+Q_{12}^{B}>0$ for
PbTiO$_{3}$. Note, that the $T_{cl}^{B}$ and $c_{B}$ values were obtained
assuming the temperature dependence of the low frequency (1 kHz) dielectric
permittivity is close to the static one. Calculation of the static
susceptibility via the maximum value of $\varepsilon _{xx}^{^{\prime \prime
}}$ measured at $T=50^{o}C$ \cite{5} has shown that the low frequency $%
\varepsilon _{xx}^{^{\prime }}$ was in fact close to the static value.

\vspace*{5mm}
\begin{figure}[th]
\centerline{\centerline{\psfig{figure=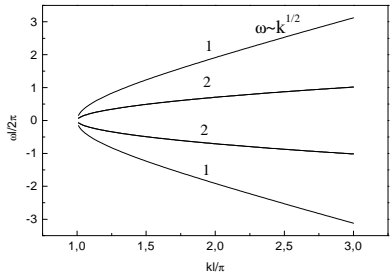,width=0.7\columnwidth}}}
\centerline{\centerline{\psfig{figure=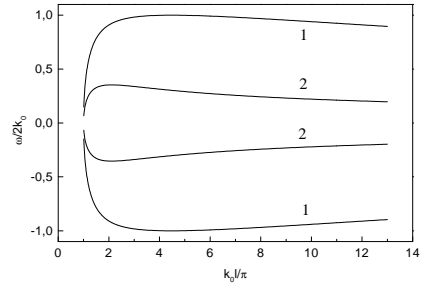,width=0.7\columnwidth}}}
\center{\text{b}}
\vspace*{5mm}
\caption{Nonlinear wave dispersion law for a multilayer in the anisotropic
case at $\protect\eta _{Bx}=0.2$, $\protect\eta _{Bz}=0.25$ (1); $\protect%
\eta _{Ax}=0.2$, $\protect\eta _{Az}=2$ (2) as a function of (a) the wave
vector and (b) the film thickness.}
\end{figure}

\section{Conclusion}

The model used for the calculations was a multilayer consisting of
alternating layers of two different ferroelectric materials for which the
interfaces between the layers were abrupt and the polarization at these
interfaces was taken to be zero. The interaction between the polarization in
different layers was neglected because of the rapid decay of this
interaction with distance \cite{15} and because of the zero polarization at
the interfaces. The production of sharp interfaces in multilayer structures
is now possible using the most recent oxide thin film deposition techniques
as has been demonstrated in \cite{7}. Our independent layer model produces
results that are in agreement with published results\cite{5,6,8} both for
multilayers and for single thin films. In addition, this model should also
be valid for bulk materials in which polarization gradients are present and
thus rendering calculations in the mean field approximation meaningless.

We have shown that the inhomogeneous polarization in the multilayers is at
the origin of the inhomogeneous static dielectric susceptibility. We have
determined the differential equation for the susceptibility and, from its
solution, we show that the conventional way of calculating the
susceptibility as the second derivative of the free energy \cite{9} is
approximately valid only for thick films with small polarization
inhomogeneity. The results related to the consideration of renormalization
of the free energy coefficients by homogeneous stresses in the film, to the
criteria of ''a/c'', ''c/c'' or ''a/a'' domains structure as well as to
thickness induced ferroelectric phase transition can be applied both to
multilayers and, after some transformations, to single thin films on
substrates. Several results are valid mainly for ferroelectric thin film
multilayers. In particular, the results obtained for the nonlinear
polarization waves and for the dispersion law exhibiting a critical wave
vector or, equivalently, a critical thickness, reflect the characteristic
peculiarities of the dynamic properties of thin film multilayers.
\vspace*{5mm}
\begin{figure}[th]
\centerline{\centerline{\psfig{figure=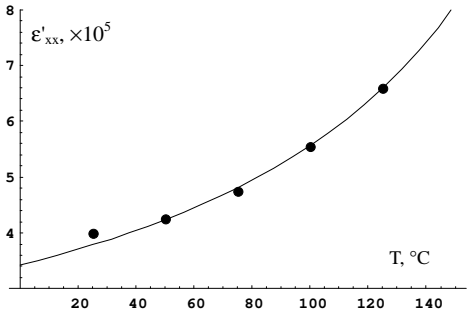,width=0.8\columnwidth}}}
\vspace*{10mm}
\caption{Temperature dependence of the dielectric permittivity of a PT-PLT
multilayer. Solid line: theory; Full circles: experiment [7].}
\end{figure}

The calculations of the inhomogeneous susceptibilities in the form of
Eq.(42) can be valid only for multilayers. Therefore the calculated
divergence of $\chi $ at the thickness induced ferroelectric phase
transition temperature may be the exclusive feature of ferroelectric thin
film multilayers. Because of the technical difficulties in producing and
studying ferroelectric thin film multilayers are more involved than for
those of single films or bulk materials, experimental results are still
somewhat restricted to just a few different series of superlattices. The
theoretical calculations of multilayer structures performed in this work are
quite complex even for the relatively simple independent layer model. This
model corresponds fairly closely to that of \cite{qu} where, for the case of
weak coupling at the interfaces, they also find the most interesting
physical phenomenon in the layers: a thickness induced ferroelectric phase
transition. Our results are in agreement with the numerical solutions in
Ref. \cite{qu}. We are now extending our calculations to obtain the static
and dynamic dielectric susceptibilities over a wide temperature region,
including the response of the ferroelectric phase for different orientations
of an external electric field. A more general model should take into account
the interfaces where gradients in the mechanical stress can appear. Thus the
calculations will increase in complexity since the equilibrium stresses must
then be calculated on the basis of the entire Euler-Lagrange equation (7b)
rather than by the simple differentiation of the free energy with respect to
stress that we have used above. Nevertheless even with the increased
computational complexity, further theoretical, as well as experimental,
investigation of ferroelectric multilayers, including those with finite
interfaces and with more than two layers in a modulation period, would seem
to be extremely desirable both for basic science and potential applications.

\section{Appendices}

\subsection{Appendix 1}

Since the free energy density $f_{B}(P_{Bz})$ and $f_{B}(P_{Bx})$ as well as
$P_{Bz}$ and $P_{Bx}$ can be obtained from one another by substitution of
their parameters (see Eq.(5a) $A\rightarrow B$ and (5b), Eq.(23) with (28a))
we shall perform a detailed calculation of $F_{B}(P_{Bz})$ and than obtain $%
F_{B}(P_{Bx})$ by this substitution.

To calculate $F_{B}(P_{Bz})$ we can take into account the first integral for
$P_{Bz}$ (the step going from Eq.(14) to Eq.(15) with $A\rightarrow B$). In
view of this first integral we can substitute the derivative for powers of $%
P_{Bz}$ in Eq.(5a) (again with $A\rightarrow B$) and obtain

\begin{equation}
f_{B}(P_{Bz})=2\beta _{33}\left( {\ \frac{dP_{Bz}}{dz}}\right) +c_{Bz}
\eqnum{A1.1a}  \label{A1.1a}
\end{equation}

\begin{equation}
c_{Bz}=b_{3}P_{Bzm}^{2}+b_{33}P_{Bzm}^{4}  \eqnum{A1.1b}  \label{A1.1b}
\end{equation}
We determine the derivative to be:

\begin{equation}
\frac{dP_{Bz}}{dz}=P_{Bzm}\sqrt{-\frac{b_{3}}{\beta _{33}}}\frac{1}{%
1+k_{Bz}^{2}}\,{\rm cn}(u_{z})\,{\rm dn}(u_{z})  \eqnum{A1.2}  \label{A1.2}
\end{equation}

\[
u_{z}=\sqrt{-\frac{b_{3}}{\beta _{33}}}\frac{z+L/2-(2j+1)l_{B}}{\sqrt{%
1+k_{Bz}^{2}}}
\]
where ${\rm cn}(u)$ and ${\rm dn}(u)$ are elliptic functions (see e.g. \cite{16}). With
respect to (A1) and (A2) the free energy $F_{B}(P_{Bz})$ can be written as

\begin{eqnarray}
&&F_{B}(P_{Bz}) =\frac{1}{L}\sum\limits_{j=0}^{N}\int%
\limits_{-L/2+(2j+1)l_{B}}^{-L/2+2(j+1)l_{B}}dz\Biggl\{c_{Bz}-  \eqnum{A1.3}
\label{A1.3} \\
&&-P_{Bzm}^{2}\frac{2b_{3}}{1+k_{Bz}^{2}}\,{\rm cn}^{2}(u_{z})\,{\rm dn}^{2}(u_{z})%
\Biggr\} = \nonumber \\
&&=\frac{l_{B}}{l_{A}+l_{B}}\Biggl\{c_{Bz}-P_{Bz}^{2}\frac{2b_{3}}{\sqrt{%
1+k_{Bz}^{2}}}\frac{1}{l_{Bz}}\times  \nonumber \\
&&\times \int\limits_{0}^{l_{Bz}/\sqrt{1+k_{Bz}^{2}}}{\rm cn}^{2}(u)\,{\rm dn}^{2}(u)%
\Biggr\}\,du.
\nonumber
\end{eqnarray}

Since $l_{Bz}=2\sqrt{1+k_{Bz}^{2}}K(k_{Bz})$ (see Eq.(25) with $A\rightarrow
B$) the last integral in (A3) has the form

\begin{equation}
I=\int\limits_{0}^{2K(k_{Bz})}{\rm cn}^{2}(u)\,{\rm dn}^{2}(u)\,du  \eqnum{A1.4}
\label{A1.4}
\end{equation}
We will now calculate the integral (A4) (see \cite{26})
\begin{eqnarray}
&&I=2\int\limits_{0}^{2K(k_{Bz})}(1-{\rm sn}^{2}u)(1-k_{Bz}{\rm sn}^{2}u)du=
\nonumber \\
&&=2\left\{ \ K(k_{Bz})-(1+k_{Bz}^{2})I_{1}+k_{Bz}^{2}I_{2}\right\},
\eqnum{A1.5}  \label{A1.5}
\end{eqnarray}

where

\[
I_{1}=\int\limits_{0}^{2K(k_{Bz})}{\rm sn}^{2}u\,du,
\]

\begin{eqnarray}
&&I_{2}=\int\limits_{0}^{2K(k_{Bz})}{\rm sn}^{4}u\;du=  \nonumber \\
&&\frac{1}{3k_{Bz}^{2}}{\rm cn}(u)\,{\rm dn}(u)\,{\rm sn}(u)\mid _{0}^{K(k_{Bz})}+  \nonumber
\\
&&\frac{2}{3}\frac{(1+k_{Bz}^{2})}{k_{Bz}^{2}} \int%
\limits_{0}^{K(k_{Bz})}{\rm sn}^{2}u\,du-\frac{1}{3k_{Bz}^{2}}K(k_{Bz}).
\eqnum{A1.6}  \label{A1.6}
\end{eqnarray}

Since ${\rm cn}(K)={\rm sn}(0)=0$, we obtain

\begin{equation}
I_{2}=-\frac{1}{3k_{Bz}^{2}}K(k_{Bz})+\frac{2}{3}\frac{(1+k_{Bz}^{2})}{%
k_{Bz}^{2}}I_{1}  \eqnum{A1.7}  \label{A1.7}
\end{equation}

We calculate $I_{1}:$
\begin{eqnarray}
&&I_{1}=\frac{1}{k_{Bz}^{2}}\left\{ u-E({\rm am}(u),k_{Bz})\right\} \mid
_{0}^{K(k_{Bz})}=  \nonumber \\
&&=\frac{1}{k_{Bz}^{2}}\left\{ K(k_{Bz})-E(\pi /2,k_{Bz})\right\}=  \nonumber
\\
&&=\frac{1}{k_{Bz}^{2}}\left\{ k_{Bz}-E(k_{Bz})\right\},  \eqnum{A1.8}
\label{A1.8}
\end{eqnarray}
where $E(k_{Bz})=\int\limits_{0}^{\pi /2}\sqrt{1-k_{Bz}^{2}\sin ^{2}\theta }%
d\theta$ is complete elliptic integral of the second kind.

Taking into consideration (A1.5), (A1.7), and (A1.8), the integral (A1.4)
can be rewritten as

\begin{equation}
I=\frac{2}{3}\left[ {\ }2K(k_{Bz})-\frac{(1+k_{Bz}^{2})}{k_{Bz}^{2}}%
(K(k_{Bz})-E(k_{Bz})\right]  \eqnum{A1.9}  \label{A1.9}
\end{equation}

Substitution of (A1.9) and (A1.1a) into (A1.3) gives

\begin{eqnarray}
&&F_{B}(P_{Bz})=\frac{l_{B}P_{Bzm}^{2}}{l_{A}+l_{B}}\Biggl\{
b_{3}+b_{33}P_{Bzm}^{2}-\nonumber \\
&&-\frac{b_{3}}{(1+k_{Bz}^{2})K(k_{Bz})}\times
\nonumber \\
&&\times\left[
2K(k_{Bz})-\frac{1+k_{Bz}^{2}}{k_{Bz}^{2}}(K(k_{Bz})-E(k_{Bz}))\right]
\Biggr\}  \eqnum{A1.10}  \label{A1.10}
\end{eqnarray}

To express (A1.10) in terms of the parameter $k_{Bz}$ and the coefficients $%
b_{3}$, $b_{33}$ we first substitute $%
P_{Bzm}^{2}/P_{Bz0}^{2}=2k_{Bz}^{2}/(1+k_{Bz}^{2})$ (see Eq.(19)) and then
we take into account that $P_{Bz0}^{2}=-b_{3}/(2b_{33})$. Thus we have

\begin{eqnarray}
&&F_{B}(P_{Bz})=-\frac{l_{B}}{l_{A}+l_{B}}\frac{b_{3}^{2}}{b_{33}}\frac{1}{%
(1+k_{Bz})^{2}}\frac{2}{3}\times \nonumber \\
&&\times \Biggl\{ \frac{k_{Bz}^{2}}{2}+1-(1+k_{Bz}^{2})%
\frac{E(k_{Bz})}{K(k_{Bz})}\Biggr \}  \eqnum{A1.11}  \label{A1.11}
\end{eqnarray}
and $F_{B}(P_{Bx})$ can be obtained from (A1.11) by substituting $k_{Bx}$
for $k_{Bz}$, $b_{1}$ and $b_{11}$ for $b_{3}$ and $b_{33}$ respectively.

This gives the inequality (32) and the function (33).

\subsection{Appendix 2}

Let's consider the solution of Eq.(37b) for the dielectric susceptibility:
\begin{equation}
\frac{d^{2}\chi (\xi )}{d\xi ^{2}}+\left( 1+m-6m\,{\rm sn}^{2}(\xi ,m)\right) \chi
(\xi )=-2(1+m)\chi _{t},  \eqnum{A2.1}  \label{A2.1}
\end{equation}
with the boundary conditions
\begin{equation}
\chi (\xi =0)=\chi (\xi =2K(m))=\chi _{s}  \eqnum{A2.2}  \label{A2.2}
\end{equation}
Here for the sake of simplicity we have omitted all subscripts and have
denoted $m=k_{Bx}^{2}$. In accordance with the theory of linear differential
equations we have to solve the homogeneous part of equation (A2.1), which is
a particular case of the Lam\'{e} equation \cite{21,22}. A solution of the
homogeneous equation is the following \cite{22}:
\begin{equation}
\chi _{1}(\xi )=C_{1}\,{\rm cn}(\xi ,m)\,{\rm dn}(\xi ,m),  \eqnum{A2.3}  \label{A2.3}
\end{equation}
A second solution can be easily obtained by varying the constant $C_{1}$ and
it has the form:
\begin{equation}
\chi _{2}(\xi )=C_{2}\;{\rm cn}(\xi ,m)\;{\rm dn}(\xi ,m)\int_{0}^{\xi }\frac{d\zeta }{%
{\rm cn}^{2}(\zeta ,m)\;{\rm dn}^{2}(\zeta ,m)}.  \eqnum{A2.4}  \label{A2.4}
\end{equation}

The particular solution of the inhomogeneous equation (A2.1) can be obtained
by varying the constants $C_1$ and $C_2$ in the sum of the solutions (A2.3)
and (A2.4). This procedure leads to:
\begin{eqnarray}
&&\chi _i(\xi )=-2(1+m)\chi _t\,{\rm cn}(\xi ,m)\,{\rm dn}(\xi ,m)\times
\nonumber \\
&&\times \int_0^\xi \frac{
{\rm sn}(\zeta ,m)\;d\zeta }{{\rm cn}^2(\zeta ,m)\,{\rm dn}^2(\zeta ,m)}.  \eqnum{A2.5}
\label{A2.5}
\end{eqnarray}

After substitution of the variable $\zeta $ by $\varphi ={\rm am}(\zeta )$, this
integral has the form $I_{1}=\int_{0}^{{\rm am}(\xi )}(1-\sin (\varphi
)^{2})^{-1}(1-m\sin (\varphi )^{2})^{-\frac{3}{2}}$. The integral $I_{1}$
can be put in standard form (see \cite{16}), and its substitution into
Eq.(A2.4) gives us the direct form of the second fundamental solution of the
equation (A2.1):
\begin{eqnarray}
\chi _{2}(\xi ) &=&\left( \xi -\frac{1+m}{1-m}E({\rm am}(\xi ),m)\right)%
{\rm cn}(\xi ,m)\,{\rm dn}(\xi ,m)+  \nonumber \\
&&+\frac{{\rm sn}(\xi ,m)\,({\rm cn}^{2}(\xi ,m)+m^{2}{\rm dn}^{2}(\xi ,m))}{1+m}.
\eqnum{A2.6}  \label{A2.6}
\end{eqnarray}

The integral in the expression (A2.5) can be reduced, by the substitution $%
t={\rm cn}(\zeta ,m)$, to that which can be found in various sources\cite{18}.Then
Eq.(A2.5) can be rewritten as
\begin{eqnarray}
&&\chi _{i}(\xi )=2\chi _{t}\frac{1+m}{\left( 1-m\right) ^{2}}\biggl( \left(
1+m\right) \,{\rm cn}(\xi ,m)\,{\rm dn}(\xi ,m)+ \nonumber \\
&&+2m\,{\rm sn}^{2}(\xi ,m)-\left( 1+m\right)\biggr)  \eqnum{A2.7}  \label{A2.7}
\end{eqnarray}
Finally the complete solution of Eq.(A2.1) has the following form:
\begin{equation}
\chi (\xi )=C_{1}\chi _{1}(\xi )+C_{2}\chi _{2}(\xi )+\chi _{i}(\xi )
\eqnum{A2.8}  \label{A2.8}
\end{equation}
where the constants $C_{1},C_{2}$ are determined from the boundary
conditions (A2.2). When using the properties of elliptic functions \cite{16}%
, the constants $C_{1},C_{2}$ can be found to have the form:
\begin{eqnarray}
&&C_{1} = \chi _{S},\eqnum{A2.9} \\
&&C_{2} = \frac{(1-m)^{2}\chi _{S}+2\chi _{t}(1+m)^{2}}{%
(1+m)E(m)-(1-m)K(m)}\left( \frac{1}{1-m}\right) .\nonumber
\end{eqnarray}
Substitution of (A2.3), (A2.6), (A2.7) and (A2.9) in (A2.8) gives us the
expression (38).

\subsection{Appendix 3}

We will calculate the susceptibility from Eq.(39d) where, for the sake of
simplicity, we have omitted all subscripts and have denoted $m=k_{Bx}^{2}$:
\begin{equation}
\chi (m)=\frac{1}{2K(m)}\int\limits_{0}^{2K(m)}\chi (\xi ,m)d\xi
\eqnum{A3.1}  \label{A3.1}
\end{equation}
using Eq.(38) for $\chi (\xi ,m)$. It is convenient to divide the integral
in the expression (A3.1) into the following three parts:
\begin{equation}
\int\limits_{0}^{2K(m)}\chi _{1}(\xi )d\xi =\int\limits_{0}^{2K(m)}{\rm cn}(\xi
)\,{\rm dn}(\xi )\,d\xi ={\rm sn}(\xi )\mid _{0}^{2K(m)}=0,  \eqnum{A3.2a}  \label{A3.2a}
\end{equation}
\begin{eqnarray}
&&\int\limits_{0}^{2K(m)}\chi _{2}(\xi )d\xi =\nonumber \\
&&=\frac{1+m}{1-m}%
\int\limits_{0}^{2K(m)}{\rm sn}(\xi )\,(1+m-2m\,{\rm sn}^{2}(\xi ))\,d\xi =  \nonumber \\
&&=-\frac{1+m}{1-m}{\rm cn}(\xi )\,{\rm dn}(\xi )\mid _{0}^{2K(m)}=2\frac{1+m}{1-m},
\eqnum{A3.2b}  \label{A3.2b}
\end{eqnarray}
\begin{eqnarray}
&&\int\limits_{0}^{2K(m)}\left( \left( 1+m\right) \left( {\rm cn}(\xi )\,{\rm dn}(\xi
)-1\right) +2m\,{\rm sn}^{2}(\xi )\right) d\xi =\nonumber \\
&&=2((1-m)K(m)-2E(m)).  \eqnum{A3.2c}
\label{A3.2c}
\end{eqnarray}
Grouping equations (A3.2a), (A3.2b) and (A3.2c), one can obtain Eq.(A3.3)
which coincides with Eq.(40):
\begin{eqnarray}
&&\chi (m)=\frac{1}{K(m)}\frac{1+m}{(1-m)^{2}}\times \nonumber \\
&&\times \Biggl( \frac{(1-m)^{2}\chi_{S}+2\chi _{t}(1+m)^{2}}{(1+m)E(m)-(1-m)K(m)}+
\nonumber \\
&&+2\chi
_{t}((1-m)K(m)-2E(m))\Biggr) .  \eqnum{A3.3}  \label{A3.3}
\end{eqnarray}

\subsection{Appendix 4}

It is necessary to consider two extreme cases of (A3.3).

\subsection{Case 1: $m\rightarrow 1$ (thick film limit).}

When we expand the elliptic functions as a series about $\mu
=(1-m)\rightarrow 0$ we obtain:
\begin{eqnarray}
K(\mu ) &=&\frac{\pi }{2}\left( 1+\frac{1}{4}\mu +\frac{9}{64}\mu
^{2}+\ldots \right)  \nonumber \\
E(\mu ) &=&\frac{\pi }{2}\left( 1-\frac{1}{4}\mu -\frac{3}{64}\mu
^{2}+\ldots \right)  \eqnum{A4.1}  \label{A4.1}
\end{eqnarray}
and using the Legendre relation \cite{16}
\[
E(m)K(1-m)+E(1-m)K(m)-K(m)K(1-m)=\frac{\pi }{2}
\]
we find that:
\begin{eqnarray}
&&(1+m)E(m)-(1-m)K(m)= \nonumber \\
&&=2-\frac{3}{2}\mu +\frac{3}{32}\mu ^{2}-\frac{3}{8}\mu
^{2}K(m)+\ldots  \eqnum{A4.2a}  \label{A4.2a}
\end{eqnarray}
and
\begin{eqnarray}
&&2E(m)-(1-m)K(m)= \nonumber \\
&&=2-\frac{1}{2}\mu -\frac{5}{32}\mu ^{2}+\frac{1}{8}\mu
^{2}K(m)+\ldots  \eqnum{A4.2b}  \label{A4.2b}
\end{eqnarray}
After substituting Eqs.(A4.2a) and (A4.2b) into the expression (A3.3) and
keeping in mind that $K(m)\rightarrow \infty $ as $m\rightarrow 1$ we find
the following limiting expression:
\begin{eqnarray}
&&\chi (m\rightarrow 1)= \nonumber \\
&&=\left( \chi _{t}+\left( \chi _{s}+\frac{3}{4}\chi
_{t}\right) \frac{1}{K(m)}\right) \left( 1-\frac{1-m}{2}\right) \rightarrow
\chi _{t} \nonumber \\
\eqnum{A4.3}  \label{A4.3}
\end{eqnarray}

\subsection{Case 2: $m\rightarrow 0$ (paraelectric thin film limit).}

For $m\rightarrow 0$ the thickness of the layer tends to the critical value $%
l_{zx}=\pi $ at which the transition to the paraelectric phase occurs$.$
Since at $m\rightarrow 0,$ $K(m)\rightarrow (\pi /2)(1+m/4)$ and$\
E(m)\rightarrow (\pi /2)(1-m/4)$ (which can be seen from the integrals of
Eqs. (24) and (A1.8)), we obtain from Eq.(A3.3):
\begin{equation}
\chi (m\rightarrow 0)\rightarrow \left( \chi _{s}+2\chi _{t}\right) \frac{8}{%
3\pi ^{2}m}  \eqnum{A4.4}  \label{A4.4}
\end{equation}
which coincides with Eq.(41).

\subsection{Appendix 5}

Let us find the approximate dependence of the susceptibility (A3.3) on
temperature. Only the dimensionless thickness $l_{Bx}$ depends on
temperature in our theory (see Eq.(29)):
\begin{equation}
l_{Bx}=\sqrt{\frac{a_{0}^{B}}{\beta _{44}}\left( T_{cx}^{B}-T\right) }l_{B}
\eqnum{A5.1}  \label{A5.1}
\end{equation}
In the vicinity of the thickness induced transition temperature $%
T=T_{cl}(l_{Bx}=\pi )$ the expression (A5.1) can rewritten as:

\begin{equation}
l_{Bx}\approx \pi \left( 1+\frac{1}{2}l_{B0}^{2}\frac{T_{cl}^{B}-T}{%
T_{cx}^{B}}\right)  \eqnum{A5.2}  \label{A5.2}
\end{equation}
where $l_{B0}$ and $T_{cl}$ are given by Eq.(30).

On the other hand it follows from the expansion of Eq.(25) at $%
m_{1}\rightarrow 0$ that:
\[
l_{Bx}=2\sqrt{1+m_{1}}K(m_{1})\approx \pi \left( 1+\frac{3}{4}m_{1}\right)
\]
\begin{equation}
m_{1}\approx \frac{4}{3}\left( \frac{l_{Bx}}{\pi }-1\right)  \eqnum{A5.3}
\label{A5.3}
\end{equation}
Substitution of Eq.(A5.2) into Eq.(A5.3) yields
\begin{equation}
m_{1}\approx \frac{2}{3}\frac{l_{B}^{2}}{l_{B0}^{2}}\frac{T_{cl}^{B}-T}{%
T_{cx}^{B}}  \eqnum{A5.4}  \label{A5.4}
\end{equation}
From this temperature dependence of $m_{1}$ we can find from Eq.(A4.4) the
susceptibility $\chi _{xx}$ as a function of temperature. In order to obtain
the temperature dependence of the susceptibility we have to choose a
specific susceptibility $\chi _{s}(T)$\ at the interface. We considered two
forms for $\chi _{s}$ on temperature and obtained the corresponding formulae
for $\chi _{xx}(T):$%
\[
\chi _{s}=const
\]
\begin{equation}
\chi _{xx}(T)\approx \left( 4\chi _{s}\frac{\beta _{44}}{l_{B}^{2}}+\frac{2}{%
\pi ^{2}}\right) \frac{1}{a_{0}^{B}\left( T_{cl}^{B}-T\right) }
\eqnum{A5.5a}  \label{A5.5a}
\end{equation}
\[
\chi _{s}=\alpha \chi _{t}=\frac{\alpha }{4a_{0}^{B}\left(
T_{cx}^{B}-T\right) }
\]
\begin{equation}
\chi _{xx}(T)\approx \frac{2+\alpha }{\pi ^{2}}\frac{1}{a_{0}^{B}\left(
T_{cl}^{B}-T\right) }  \eqnum{A5.5b}  \label{A5.5b}
\end{equation}
Under the assumption $\chi _{s}=0$, Eqs.(A5.5a) and (A5.5b) lead us to
Eq.(43).

\bigskip

The authors are grateful to L.\ Lahoche for worthwhile discussions. One of
us (M.\ D.\ G.) would like to thank the University of Picardy for financial
support.


\begin{references}
\bibitem{1}  H. T. Grahn, Semiconductor superlattices: Growth and Electronic
Properties (World Scientific, Singapore, 1995).

\bibitem{2}  T. Shinjo and T. Takada, Metallic Superlattices: Artificially
Structured Materials (Elsevier, New York, 1987).

\bibitem{3}  M. G. Cottam, Linear and Nonlinear Spin Waves in Magnetic Films
and Superlattices (World Scientific, Singapore, 1994).

\bibitem{4}  I. Bozovic, Superconducting Superlattices and Multilayers
(SPIE, Bellingam, 1994)

\bibitem{sw}  S. L. Swartz and V. E. Wood, Condensed Matter News, {\bf 1}, N
5, 4 (1992).

\bibitem{tab}  H. Tabata and T. Kawai, Appl. Phys. Lett., {\bf 70}, 321
(1997).

\bibitem{5}  Y. Kim, R. A. Gerhardt and A. Erbil, Phys. Rev. B, {\bf 55},
8766 (1997).

\bibitem{6}  E. D. Specht, H.-M. Christen, D. P. Norton and L. A. Boatner,
Phys. Rev. Lett., {\bf 80}, 4317 (1998).

\bibitem{7}  J. C. Jiang, X. Q. Pan, W. Tian, C.D.\ Theis and D.\ Schlom,
Appl. Phys. Lett., {\bf 74}, 2851 (1999).

\bibitem{8}  F. Le Marrec, R. Farhi, M. El. Marssi, J. L. Dellis, M. G.
Karkut and D.\ Ariosa (Phys. Rev. B, to be published).

\bibitem{9}  Bai Shaoping Li, J. A. Eastman, J. M. Vetrone, R. E. Newnham
and L. E. Cross, Phil. Mag. B, {\bf 76}, 47 (1997).

\bibitem{qu}  B.\ D.\ Qu, W.\ L.\ Zhong and R.\ H.\ Prince, Phys.\ Rev.\ B,
{\bf 55},{\bf \ }11218 (1997).

\bibitem{10}  See, for example, J.\ Mathews and J.\ L.\ Walker, Mathematical
methods of physics (W.\ A.\ Benjamin, New York, 1965).

\bibitem{11}  W. D. Nix, Metallurgical Transactions, {\bf 20A}, 2217 (1989).

\bibitem{12}  G. A. Rossetti, Jr., L. E. Cross and K. Kushida, Appl. Phys.
Lett., {\bf 59}, 2524 (1991).

\bibitem{13}  N. A. Pertsev, A. G. Zembilgotov and A. K. Tagantsev, Phys.
Rev. Lett., {\bf 80}, 1988 (1998).

\bibitem{14}  J. S. Speck and W. Pompe, J. Appl. Phys., {\bf 76}, 466 (1994).

\bibitem{15}  D. R. Tilley, Ferroelectric Thin Films (Gordon and Breach,
Amsterdam, 1996) p.11.

\bibitem{16}  M. Abramowitz and A. Stegun, Handbook of Matematical Functions
(Dover Publications Inc., N.Y.).

\bibitem{17}  E. Yahnke, F. Emde and P.Losch, Tables of Higher Functions
(McGraw-Hill, N. Y., 1960).

\bibitem{18}  I. S. Gradshteyn and I. M. Ryzhik, Tables of Integrals, Series
and Products (Academic Press, New York, 1965).

\bibitem{19}  Y. Ishibashi, H. Orihara and D. R. Tilley, J. Phys. Soc. Jap.,
{\bf 67}, 3292 (1998).

\bibitem{20}  Y. G. Wang, W. L. Zhong and P. L. Zhang, Phys. Rev. B, {\bf 51}%
, 5311 (1995).

\bibitem{21}  E. Whittaker and G. Watson, A course of modern analysis
(Cambridge University Press, 1958).

\bibitem{22}  A. Erdelyi, Higher Transcendental Functions (McGraw Hill, N.\
Y., 1953).

\bibitem{23}  V. G. Vaks, Vvedenie v microscopicheskuyu teoriyu
segnetoelectrikov (Moscow, Nauka, 1973).

\bibitem{24}  see for example V. L. Ginzburg, JETF, {\bf 19}, 36 (1949);
UFN, {\bf 38}, 490 (1949); FTT, {\bf 2}, 2031 (1960).

\bibitem{25}  see for example A. M. Kosevich and A. S. Kovalev, Introduction
to nonlinear physical mechanics (Naukova Dumka, Kiev, 1989).

\bibitem{26}  A. P. Prudnikov, Yu. A. Bruchkov and O. I. Marichev, Integrals
and Series, {\bf 3}, (Moscow, Nauka, 1986).
\end{references}
\end{document}